\documentclass[twocolumn,usenames,dvipsnames]{aastex62}

\usepackage{amsmath}
\usepackage{apjfonts}
\usepackage{amssymb}
\usepackage{xspace}
\usepackage{hyperref}
\usepackage{natbib}
\usepackage{multirow}
\usepackage{graphicx}
\usepackage{epstopdf}
\usepackage{epsf}
\usepackage{subfigure}
\usepackage{xcolor}
\usepackage{rotating}
\usepackage{mathrsfs}
\usepackage{slashbox}
\usepackage{multirow}
\usepackage{lipsum}
\usepackage[normalem]{ulem}

\def\Msun{M_\odot}

\def\xHI{{x_{\rm HI}}}

\begin{document}

\title{Recovering Density Fields inside Quasar Proximity Zones at $z\sim 6$}

\author{Huanqing Chen}
\affiliation{Department of Astronomy \& Astrophysics; 
The University of Chicago; 
Chicago, IL 60637, USA}

\author{Nickolay Y.\ Gnedin}
\affiliation{Theoretical Physics Department; 
Fermi National Accelerator Laboratory;
Batavia, IL 60510, USA}
\affiliation{Kavli Institute for Cosmological Physics;
The University of Chicago;
Chicago, IL 60637, USA}
\affiliation{Department of Astronomy \& Astrophysics; 
The University of Chicago; 
Chicago, IL 60637, USA}

\correspondingauthor{Huanqing Chen}
\email{hqchen@uchicago.edu}

\begin{abstract}

 The matter density field at $z\sim 6$ is very challenging to probe. One of the traditional probes of the low density IGM that works successfully at lower redshifts is the Lyman-alpha forest in quasar spectra. However, at the end of reionization,  the residual neutral hydrogen usually creates saturated absorption, thus much of the information about the gas density is lost. Luckily, in a quasar proximity zone, the ionizing radiation is exceptionally intense, thus creating a large region with non-zero transmitted flux. In this study we use the synthetic spectra from simulations to investigate how to recover the density fluctuations inside the quasar proximity zones.  We show that, under ideal conditions, the density can be recovered accurately with a small scatter. We also discuss how systematics such as the quasar continuum fitting and reionization models affect the results. This study shows that by analyzing the absorption features inside quasar proximity zones we can potentially constrain quasar properties and the environments they reside in.

\end{abstract}

\section{Introduction}

The intergalactic medium (IGM) contains most of baryons of the Universe \citep[e.g.,][]{fukugita1998}. At redshifts $z=2\sim5$ the hydrogen in the Universe is mostly ionized, and the IGM can be effectively probed by the Ly$\alpha$ forest in quasar absorption spectra. Statistical tools such as the flux power spectra and the transmission probability distribution function have been developed to exploit the Ly$\alpha$ forest over the past few decades \citep{rauch1997,croft1998,mcdonald2000}. The study of the Ly$\alpha$ spectra not only helps constrain key cosmological parameters \citep{viel2005,seljak2005}, but also provides crucial information about galaxy formation, such as the thermal history of the IGM and evolution of the cosmic ionizing background \citep[see][for a recent review]{mcquinn2016}.

The quasar absorption spectra have been a very sensitive and successful tool to probe the mostly ionized ($\xHI\lesssim 10^{-5}$) universe, thanks to the exceptionally large cross-section of the Ly$\alpha$ line. However, the amount of information encoded in them appears to diminish toward higher redshifts ($z\sim 6$).  In fact, a typical quasar spectrum at $z\approx6$ shows complete absorption or "dark" gaps, also called “Gunn-Peterson troughs”. As these dark gaps contain no transmitted flux, information about the underlying distributions of cosmic gas density, ionization fraction and temperature is lost.

Although most of Ly$\alpha$ absorption at $z\sim 6$ is saturated, there are a few exceptions. A luminous quasar dominates the radiation field in its vicinity, lowering the values of the neutral fraction. As a result, in the quasar spectrum the spectral region immediately blueward of the quasar Ly$\alpha$ emission line shows some transmitted flux. Such a region is called a quasar ``proximity zone'' or ``near zone'', and can extend up to $\sim 10$ pMpc for bright quasars.

Current research on quasar proximity zones primarily focuses on exploring the sizes of the proximity zones, especially their evolution with redshift and quasar age, with the goal of constraining reionization history or the quasar lifetime \citep{fan2006,carilli2010,eilers2017}. Over the last two decades many useful insights have been gained. For example, hydrodynamic simulations predict the median proximity zone sizes very well for a reasonably long quasar age, but struggle to reproduce the observed number of exceptionally small proximity zones \citep{keating2015,eilers2017,chen2021}, implying that quasar lifetimes are perhaps short. Simulations also predict a shallow proximity zone size evolution with redshift, but, unfortunately, this may tell little about the global reionization history. This is because the traditional definition of the quasar proximity zone size\footnote{Measured to the first point where the transmitted spectra, smoothed by  $20$ \AA \  boxcar, drops under $10\%$.} underestimates the actual size of the region where the quasar radiation dominates over the (local) cosmic background and, hence, ionization state of the gas at the edge of the conventionally defined proximity zone still reflects the quasar radiation rather than the cosmic radiation background \citep{bolton2007}.

However, any coin has two sides. Since inside the proximity zone the radiation field is dominated by the quasar, the radiation profile is well approximated by simple $r^{-2}$, which makes the analysis of absorption features relatively easy. With the radiation profile known, it is straightforward to recover the gas density from the absorption spectra.
So far, there have been dozens of $z\gtrsim 6$ quasars observed with high spectral resolution, and we expect many more from future thirty-meter-class telescopes. These spectra show complex absorption features inside the proximity zones. These features have not yet been explored for the potentially rich information encoded in them. 

In this study we explore the possibility to recover density fluctuations at $z\sim 6$ with Ly$\alpha$ absorption inside quasar proximity zones. This paper is organized as follows: in \S\ 2 we describe the related equations and the simulation we use. In \S\ 3 we showcase one example to demonstrate the basic step in the density recovery procedure, and quantify the accuracy of the recovered density.
In \S\ 4 we discuss additional complexities affecting real-life observations and quantify the systematic errors one may encounter. In \S\ 5 we summarize our results and give an outlook for future applications.

\section{Methodology}

\subsection{Basic Equations}

In this subsection we describe the basic idea of recovering the density profile in a quasar proximity zone. The process is based on two key assumptions about the IGM in the proximity zone: (1) the IGM is optically thin to ionizing radiation, and (2) the IGM is in ionization equilibrium. In the next subsection, we will show that for quasars with light-bulb light curve, these two assumptions are true for the majority of sightlines.

Let us consider a toy case of gas with no peculiar velocity and an infinitely sharp Ly$\alpha$ with no thermal broadening. Similarly to the case of lower redshift Ly$\alpha$ forest, in the proximity zone the contribution $d\tau(R)$ to the optical depth at the distance $R$ from the quasar by the gas element of size $dr$ located at the distance $r$ is
\begin{equation}
\begin{split}
    d\tau(R)&=\sigma\left(\Delta\nu (R-r)\right)n_{\rm HI}(r)dr \\
    &=\frac{\pi e^2 f}{m_e c}\phi\left(\frac{\nu_0 H (r-R)}{c}\right)n_{\rm HI}(r)dr \\
    &\approx \frac{\pi e^2 f}{m_e c}\delta\left(\frac{\nu_0 H (r-R)}{c}\right)n_{\rm HI}(r)dr, \\
\end{split}
\end{equation}
where the observed frequency $\nu$ at the physical distance $R$ from the quasar is 
\[
    \nu = \nu_{\rm Ly\alpha}\left(1+z+\frac{HR}{c}\right).
\]
 The cross-section as a function of $\delta \nu$ from the line center, $\sigma (\Delta \nu)$, is related to electric charge $e$, oscillation strength of the $Ly\alpha$ line $f$, electron mass $m_e$ and the speed of light $c$, and the line profile $\phi(\Delta \nu) \approx \delta (\Delta \nu)$ for our assumed toy case of no thermal broadening. To obtain the optical depth at $R$, we need to integrate the contribution from all gas elements:
\begin{equation}
\begin{split}
    \tau(R) &=\int_0^\infty d\tau(R) = \int_0^\infty \sigma(\Delta\nu (R-r))n_{\rm HI}(r)dr\\
    &\approx \frac{\pi e^2 f}{m_e c}\frac{c}{\nu_0 H}n_{\rm HI}(R) \\
\end{split}
\end{equation}

For a bright quasar, the ionization equilibrium timescale is very short within a few physical megaparsecs (pMpc). For example, for a quasar with the ionization rate of $\approx 10^{57} \rm s^{-1}$, the ionization equilibrium timescale at $4$ pMpc is smaller than $0.1$ Myr. Therefore, it is safe to assume that
$$
\Gamma n_{\rm HI}=\alpha(T) n_{\rm HII} n_e,
$$
where $n_{\rm HI}, n_{\rm HII}$, and $n_e$ are the density of HI, HII and electrons, respectively. The factor $\Gamma$ is the photo-ionization rate and $\alpha(T)$ is the recombination rate, which is a function of gas temperature. In a typical quasar proximity zone where the radiation is dominated by the quasar and the IGM is optically thin, $\Gamma$ can be assumed to be simply inversely proportional to the square of the distance from the quasar, $\Gamma(R)=\Gamma_1 R_1^{-2}$, where $\Gamma_1$ is the photo-ionization rate at $1$ pMpc and $R_1$ is distance in pMpc. Therefore, the optical depth at $R_1$ is
\begin{equation}
\begin{split}
    {\tau}(R) &\approx \frac{\pi e^2 f}{m_e c}\frac{c}{\nu_0 H}n_{\rm HI}(R) \\
    &= \frac{\pi e^2 f}{m_e c}\frac{c }{\nu_0 H \Gamma_1} R_1^2 \alpha(T) \left(\frac{n_e}{n_{\rm H}}\right) n^2_{\rm H}
\end{split}\label{eq:alphaT}
\end{equation}
and we assumed that hydrogen is highly ionized, $n_{\rm HII}\approx n_{\rm H}$.
The first two terms are constant for a given quasar, and the recombination rate is only a weak function of temperature $\alpha\propto T^{-0.7}$. Assuming that $n_e/n_{\rm H}$ is also constant in a mostly ionized universe\footnote{Strictly speaking, this ratio would has a small jump (<$10\%$) across the HeII I-front, which is considered in Section \ref{sec: age}.}, we can compare the opacity in the proximity zone to the opacity of an imaginary sightline with the uniform density:
\begin{equation}\label{Eq: tau}
\frac{\tau(R)}{\tilde{\tau}(R)}=\frac{\alpha(T(R))}{\alpha(\tilde{T}(R))}\left(\frac{n_H(R)}{\tilde{n}_H(R)}\right)^2
\end{equation}
where tilded quantities are for the imaginary sightline with the uniform density of the mean value of the universe. Right after reionization, the temperature-density relation $T\approx n^{\gamma-1}$ has a much lower index than at local universe. Our simulation shows that $\gamma\sim 1.3$ at $z\sim 6$ and is even slightly lower after the IGM is photo-heated by the quasar. Therefore, the recombination rate $\alpha \sim n_H^{-0.2}$, which is very flat. To the first order, we can ignore this term so that
\begin{equation}\label{Eq: Delta_g}
\Delta_g \equiv \frac{n_H(R)}{\tilde{n}_H(R)} \approx\sqrt{\frac{{\tau}_{\rm Ly \alpha}}{\bar{\tau}_{\rm Ly \alpha}}}.
\end{equation}
This simple expression is the key to our density recovery procedure.

In reality, several factors can introduce deviations from this simple equation. { For example, the gas inside the proximity zone is at  about $\sim 2\times 10^4$ K,  so the line profile is a Gaussian with the Doppler parameter $b \equiv \sqrt{2}\sigma=\sqrt{2kT/m_p}\approx 18 ~\rm~ km/s$. This $\sigma$ is equivalent to $\sim 20$ pkpc at $z\sim 6$ or $\sim 0.05$\AA \ in the rest frame. Therefore, any density fluctuation smaller than $\sim 20 \rm pkpc$ is smoothed out in the spectrum.} A much more serious complication is the peculiar velocity of the gas, which can easily be $\sim 100 ~\rm~ km/s$. These complications are addressed below.

\subsection{Simulation}

In this study we quantify the accuracy of the density recovering process with synthetic spectra from simulations. The dataset we use are detailed in \citet{chen2021}. Here we briefly describe them and refer the readers to the original paper for more information.

We obtain the synthetic spectra by post-processing sightlines from one of the Cosmic Reionization on Computers (CROC) simulations. The CROC simulation does not explicitly model quasars. We draw $6930$ sightlines from the $63$ most massive halos ($M_h>1.5\times 10^{11} M_\odot$) in the simulation at $z=6.11$. These sightlines are post-processed by a 1D radiative transfer code, in which a quasar of constant luminosity shines for $30$ Myr. The quasar has a spectral index of $-1.5$ ($L_\nu\propto \nu ^ {-1.5}$) and a total ionizing photon rate 
$$ \dot{N} = \int_{\nu_{\rm HI}}^{\infty} \dot{N}_\nu d \nu = 1\times 10^{57} ~\rm~ s^{-1}.$$
This corresponds to a magnitude of $M_{1450}=-26.67$ if assuming the same spectra index from $1450$ \AA \ to $912$ \AA \  with no break.

We have gained key insights on the quasar radiation profiles from this set of simulations: the radiation profile within a few pMpc is entirely dominated by the quasar, and for the majority of the sightlines the radiation profile approaches perfect $r^{-2}$ within $0.3$ Myr.

\begin{figure}
    \centering
        \includegraphics[width=0.45\textwidth]{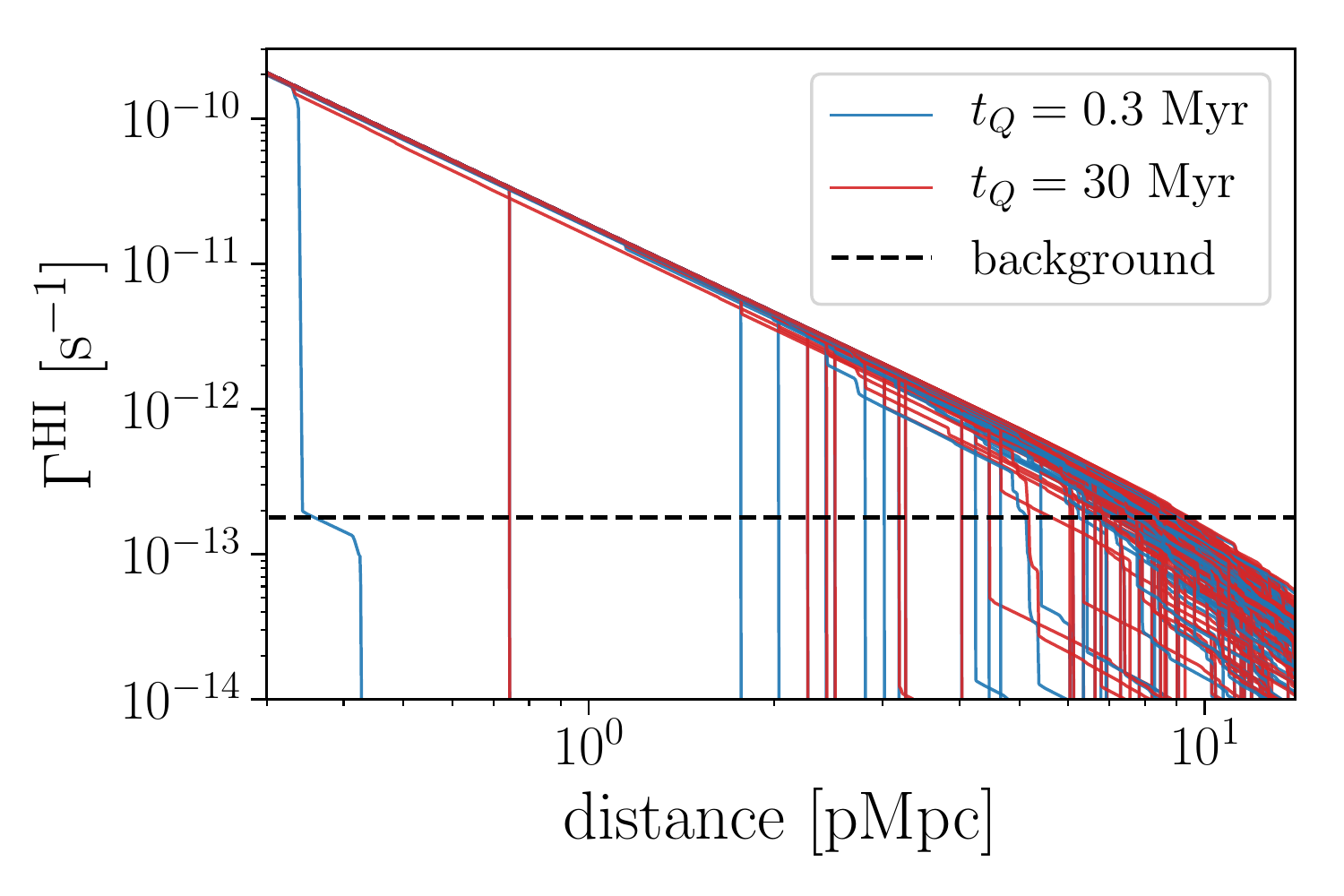}
    \includegraphics[width=0.45\textwidth]{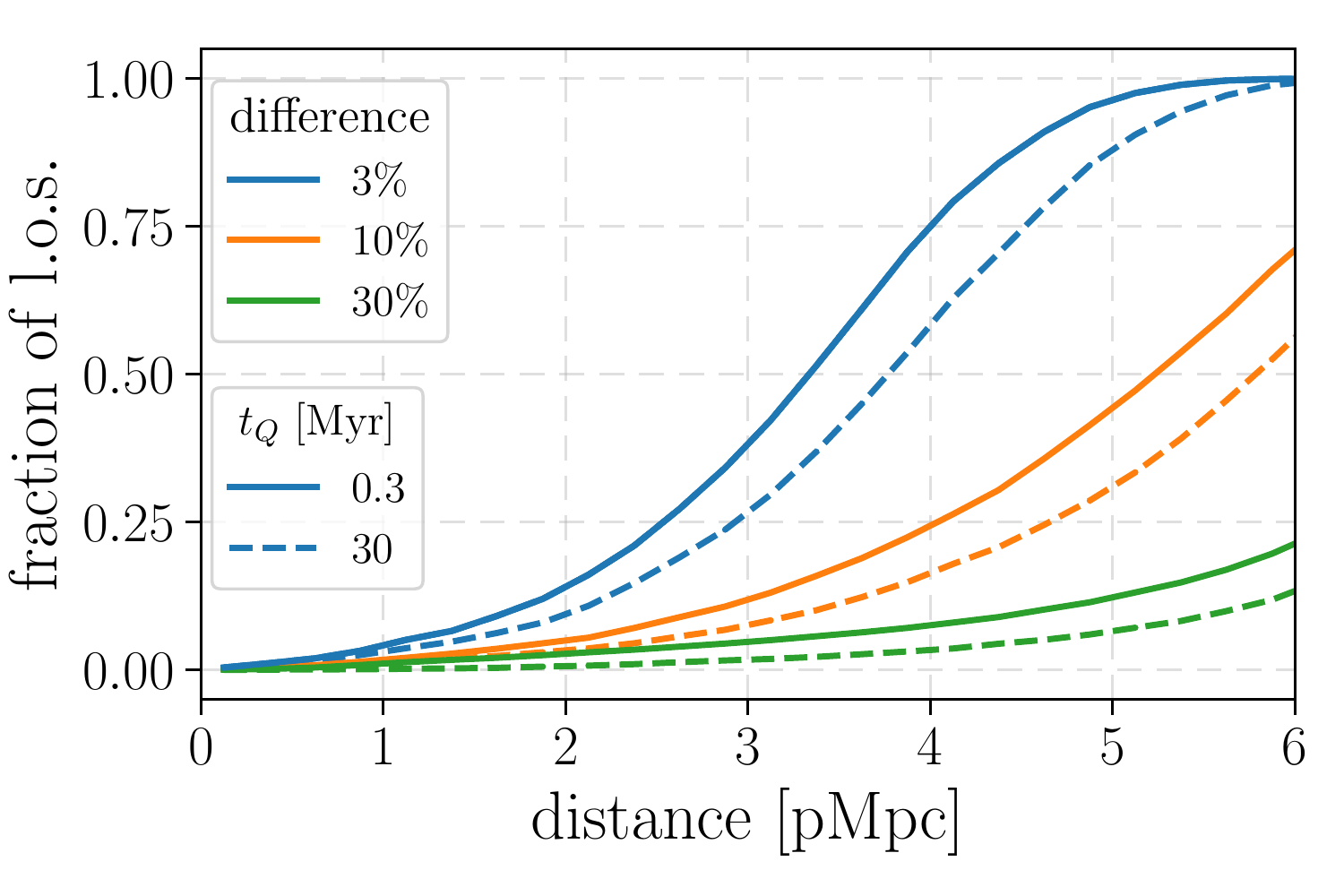}
    \caption{Upper panel: Quasar radiation profiles of $100$ randomly selected sightlines. Blue lines are for $t_Q=0.3$ Myr, while red lines are for $t_Q=30$ Myr for the same sightlines. The black dash line is the median level of the cosmic radiation background.  Lower panel: Fraction of sightlines in our sample (with $R_{\rm phy}>6$ pMpc) whose radiation profile differ from the perfect transparent IGM case by $3\%$,$10\%$, and $30\%$ at a given distance from the quasar. Solid lines are for $t_Q=0.3$ Myr and dash lines are for $t_Q=30$ Myr.}
    \label{fig:attenuationFrac}
\end{figure}

In the top panel of Figure \ref{fig:attenuationFrac} we show the quasar ionizing rate profile for $100$ random sightlines. The vast majority of them do not deviate from a perfect power-law shape of index $-2$ (fully transparent IGM) by more than 30\% within $\sim 4$ pMpc. The deviations are caused by Lyman limit systems (LLSs) or damped Lyman-$alpha$ (DLA) systems, which correspond to the sudden drops in $\Gamma^{\rm HI}$. The dashed black line show the median of the cosmic radiation background, which is an order of magnitude lower than the radiation from quasar within $4$ pMpc. 

\begin{figure}
    \centering
    \includegraphics[width=0.5\textwidth]{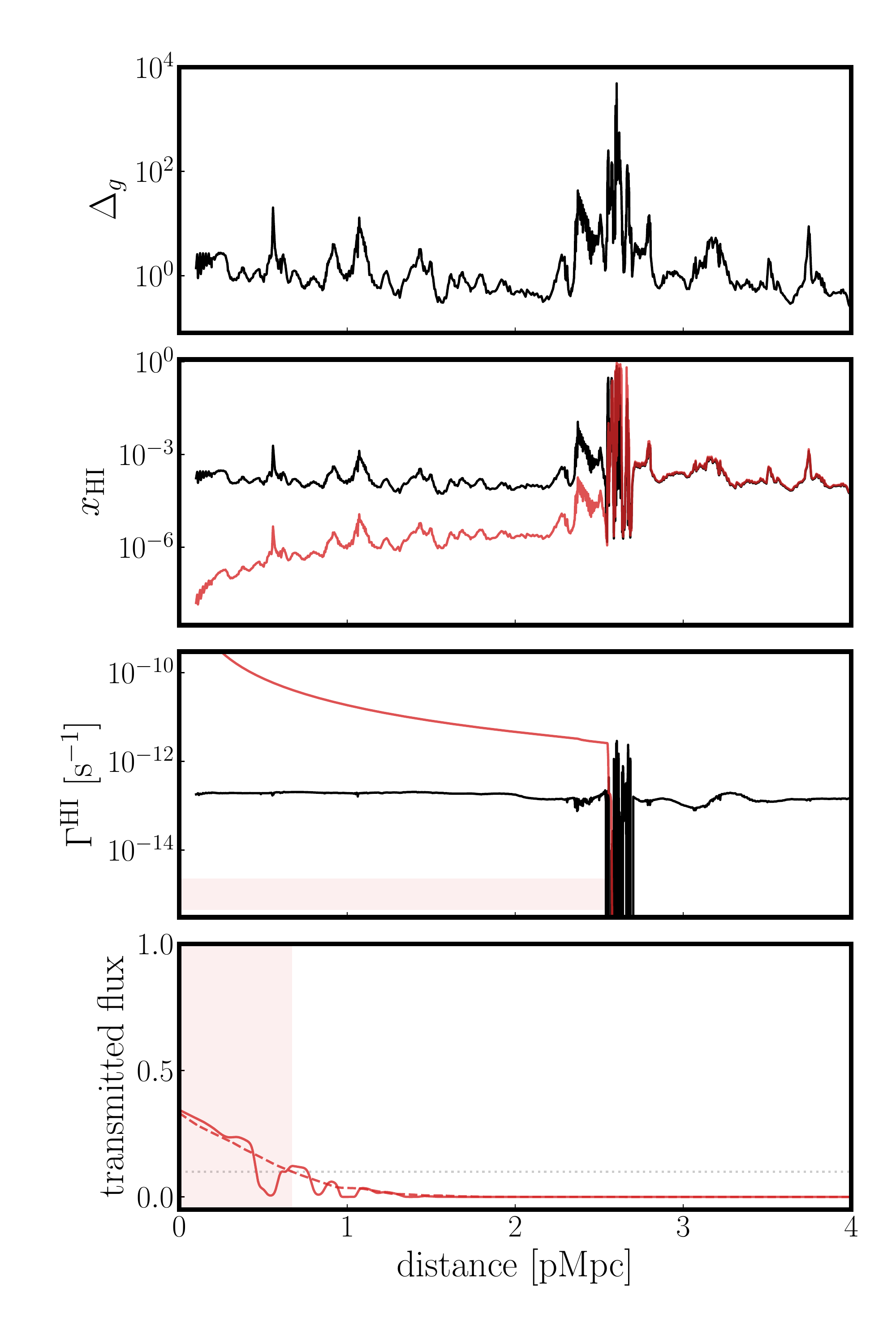}
    \caption{An example of a sightline that travels through a galaxy proximity zone at $d\sim 2.5$ pMpc.}
    \label{fig:galPZ}
\end{figure}

Note that most of the sightlines with LLSs/DLAs close to the quasar can be identified in observations since they display no transmitted flux outside the LLSs/DLAs or exhibit damping wings. Because LLSs/DLAs may contaminate other pixels, for our purpose of recovering density field we do not want to include them in our sample. We exclude them by simply removing the sightlines with physical proximity zone size (where $\Gamma^{\rm HI}_{\rm QSO}>\Gamma^{\rm HI}_{\rm bkg}$) smaller than $6$ pMpc at $t_Q=30$ Myr, so that  sightlines resembling Figure \ref{fig:galPZ} are not included due to ``absorption contamination" from the damping wing. This results in the final sample of $6001$ sightlines for this study.

One may worry about the possibility that some sightlines pass very close to foreground galaxies, where such ``galaxy proximity zones" can increase the background radiation, making the assumption of the perfect power-law $\Gamma^{\rm HI}$ invalid. We randomly chose $\sim 200$ sightlines and visually examine them and find only one of them where the radiation background fluctuates up by an order of magnitude. This fluctuation actually corresponds to a DLA, which manifests itself as the suppression in the transmitted flux due to the damping wing, rather than as an increment due to the enhanced galaxy radiation.
Such sightlines should be easy to identify observationally. Furthermore, we also randomly draw sightlines within $\sim 10$ pkpc from massive halos to check their radiation backgrounds and density fields, and we find that although the radiation background can fluctuate up by an order of magnitude on a spatial scale of $<100$ pkpc, they inevitably hit gas that are at least $100$, and most commonly $>1000$ higher than the mean gas density. Therefore these ``galaxy proximity zones"  usually correspond to LLS/DLAs.

For the majority of the sightlines the IGM within $4$ pMpc from the quasar reaches the ionization equilibrium within $1$ Myr ($\sim 1/\Gamma^{\rm HI}$). Therefore, if the quasar maintains the luminosity for longer than $\sim 1$ Myr, the ionization equilibrium assumption from the previous subsection is valid.

\section{Results}

\subsection{Recovering sightlines without peculiar velocity}

\begin{figure*}
    \centering
    \includegraphics[width=\textwidth]{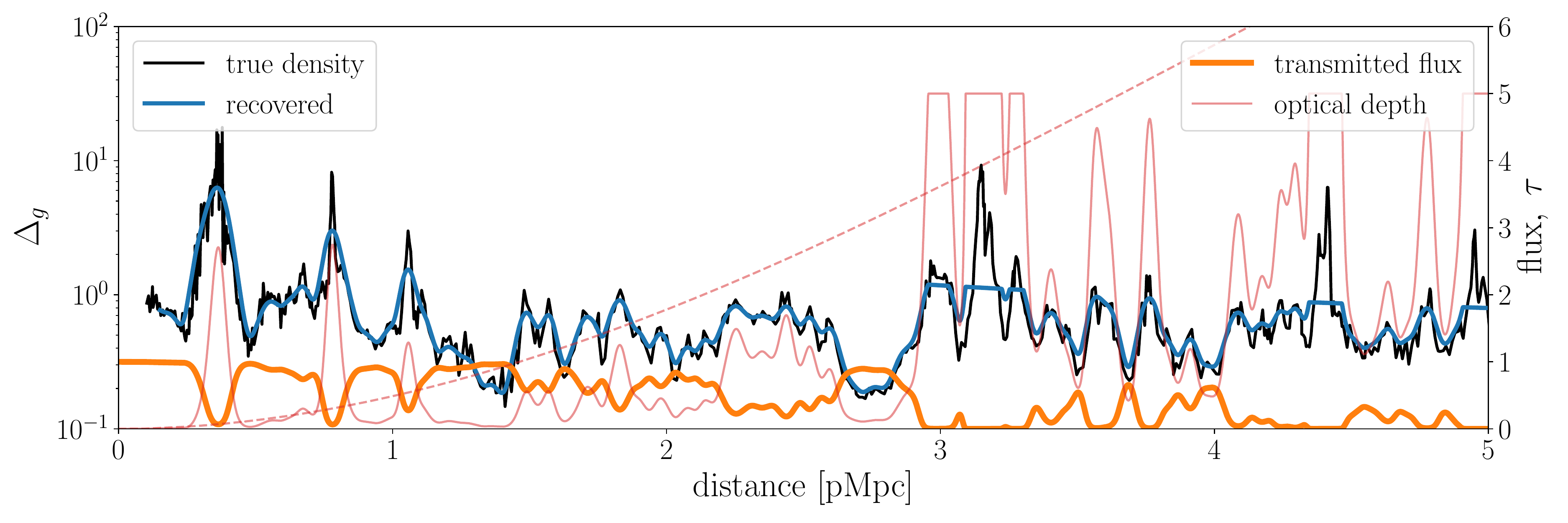}

    \caption{An example of a sightline without peculiar velocity at $z=6.11$. The magnitude of the quasar is M1450=-26.67. Its proximity zone size is above the median value, and in the distance range plotted, the quasar dominates the radiation field. The red dashed line is the optical depth of the imaginary sightline with uniform mean density of the universe (baseline).
\label{fig:novpec}
    }
    
\end{figure*}

\begin{figure*}
    \centering
    \includegraphics[width=0.4\textwidth]{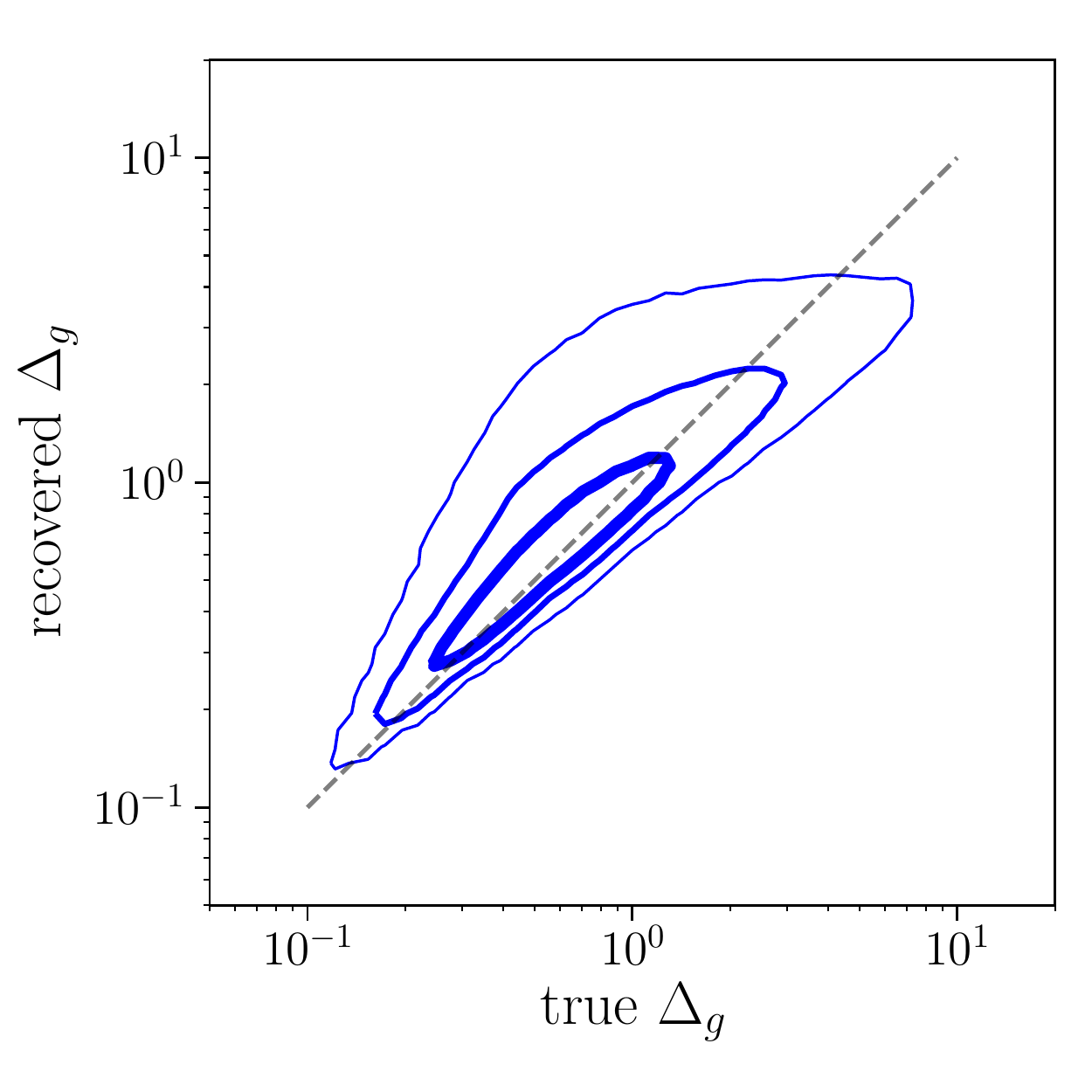}
    \includegraphics[width=0.4\textwidth]{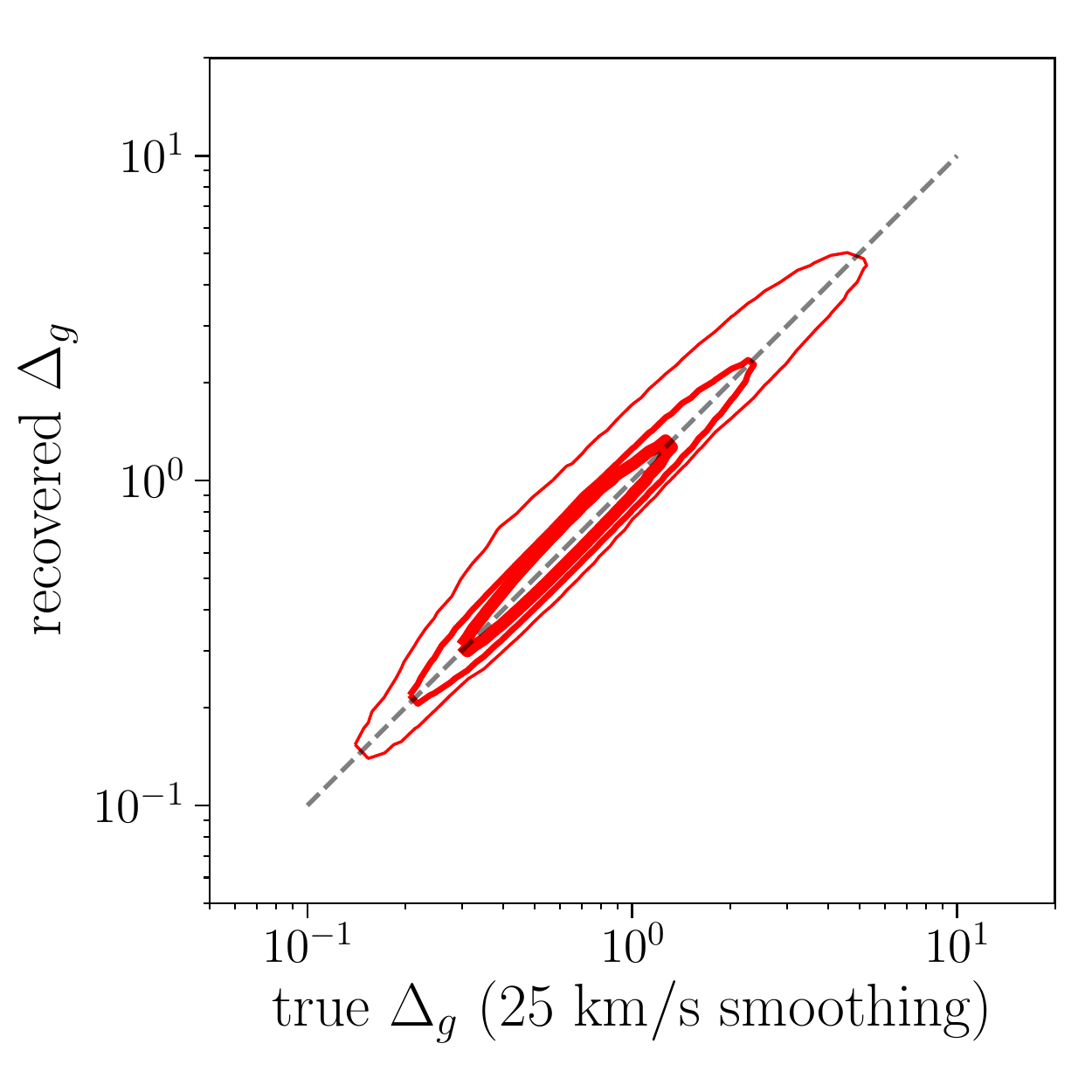}
    \caption{Scatter plot for the recovered gas density from synthetic spectra without peculiar velocity (y-axes), versus the true gas density (x-axes) for all unsaturated pixels in the distance range of $0.5-4$ pMpc from all the 6001 quasars in our sample. The true density in the left panel is not smoothed while in the right panel is smoothed by a Gaussian filter with  $b= 25$ km/s .}
    \label{fig:rspace_scatter}
\end{figure*}

\begin{figure*}
    \centering
    \includegraphics[width=0.4\textwidth]{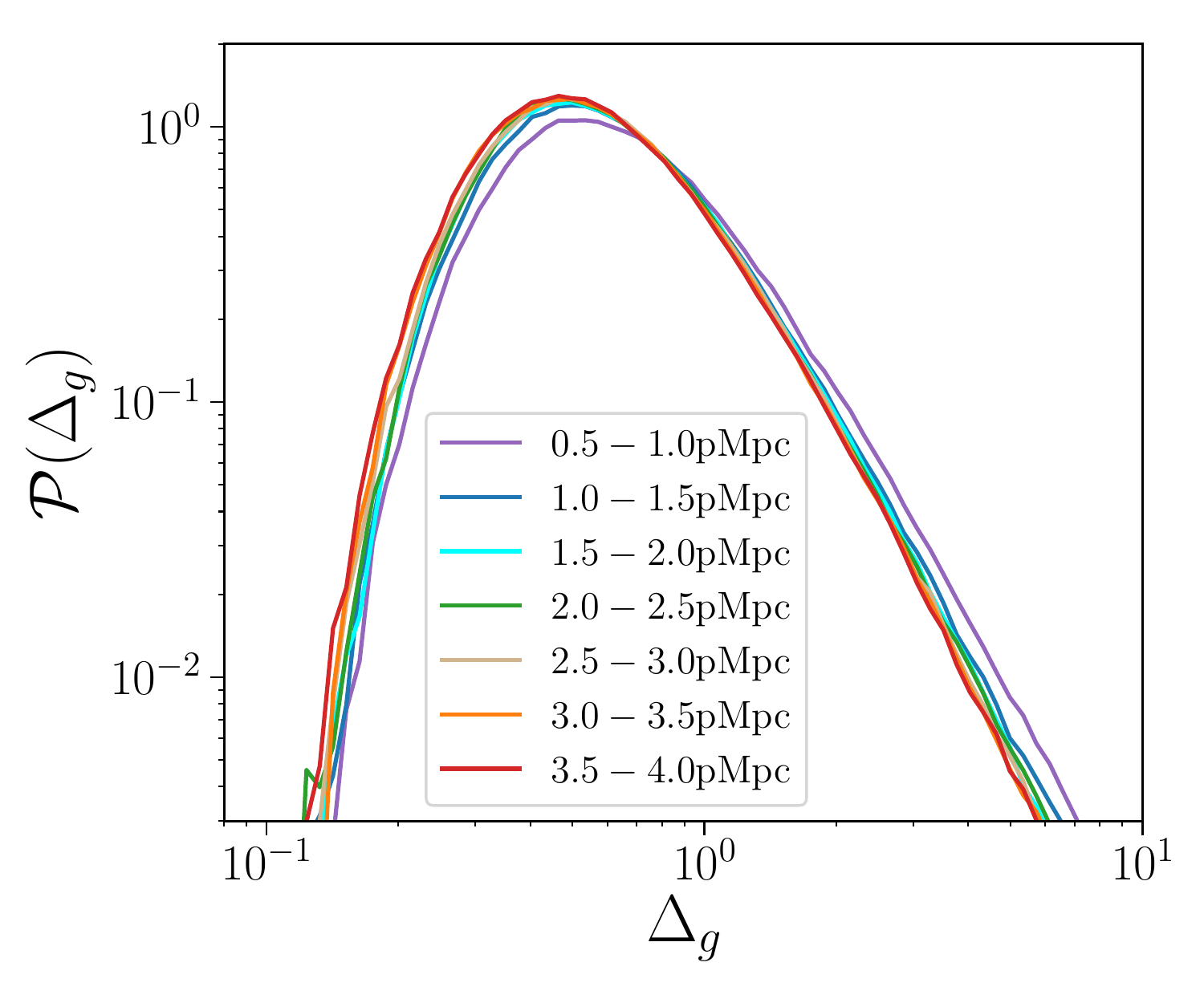}
    \includegraphics[width=0.4\textwidth]{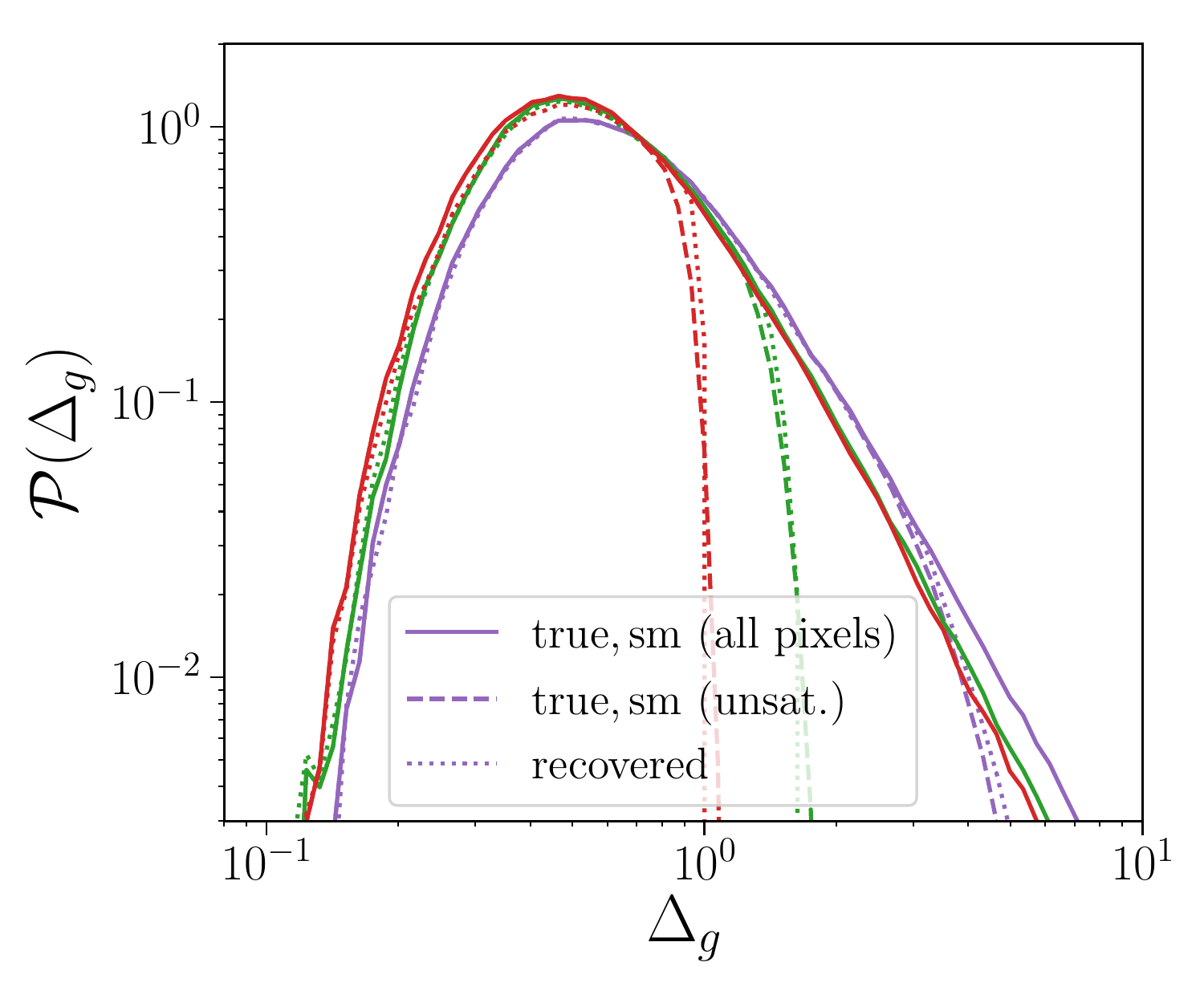}
    \caption{Density PDF in real space. Left: PDFs of the real space density, smoothed by $b=25 \rm km/s$. Different colors represent spectral regions at different distances from the quasar. These true density PDFs include all pixels in spectra, saturated and unsaturated. Right: solid curves are the same as in the left panel, while the dashed lines are true smoothed density PDFs that include only unsaturated pixels. Dotted lines are density PDFs recovered from the systhentic spectra without peculiar velocity using Equation (\ref{Eq: Delta_g}).}
    \label{fig:rspace_PDF}
\end{figure*}

We demonstrate the procedure of recovering the gas density field from our synthetic spectra. In this subsection we ignore the peculiar velocity of the gas as a illustrative toy model. 

In Figure \ref{fig:novpec} we show an example of a sightline drawn from the simulation with peculiar velocity set to zero, so that the absorption features line up exactly with the density features. The black line shows the gas density in units of the mean gas density of the universe (also called ``the density contrast'') $\Delta_g$ along the sightline. The orange line shows the transmitted flux after the quasar has been shining for $30$ Myr. The red line shows the corresponding optical depth. Because in reality we may not measure anything above $\tau >5$ (transmitted flux $\lesssim 1\%$), we set an optical depth cap of ${\rm max}(\tau )=5$, above which the optical depth is not known.

To recover the gas density contrast $\Delta_g$ using Equation (\ref{Eq: Delta_g}) we need to know the mean optical depth $\bar{\tau}_{\rm Ly\alpha}$ (we hereafter call it baseline optical depth). We obtain this also by post-processing a sightline with unifrom density fixed at the value of the cosmic mean density and initial temperature and the ionization fraction of roughly the mean values in the simulation at $z=6.11$. The exact values we adopt are $x_{\rm HI}=10^{-4}$, $x_{\rm HeI}=10^{-4}$, $x_{\rm HeII}=0.9$, $T=10^4$ K. The sightline is post-processed by the same 1D RT code for a quasar age $t_Q=30$ Myr. Note that using a different baseline does affect the accuracy of the recovered gas density, we return to that point in Section \ref{sec:discussion}.
Because of the quadratic drop of the radiation field with distance, the baseline optical depth rises quickly and exceeds $5$ at around $4$ pMpc for a $\dot{N}=1 \times 10^{57} ~\rm~ s^{-1}$ quasar. As a result, most spectra of quasars with this luminosity exhibit little transmitted flux outside that distance.

The recovered gas density $\sqrt{\tau/\bar{\tau}}$ is plotted as the blue line in Figure \ref{fig:novpec}. Because we limit the optical depth, there is a maximum gas density that we can recover. This max($\Delta_g$) depends on the distance as max($\Delta_g$)$\propto 1/R$. Therefore, outside the radius where $\tau_{\rm bl}={\rm max} (\tau)=5$, which is $\approx 3.7$ pMpc for this quasar luminosity, we can only recover gas under-density, while closer to the quasar we can recover densities up to $\Delta_g \sim 10$.
Comparing the recovered gas density (blue line) and the true gas density (black line), we find that in unsaturated regions the density is recovered very well. If we take a closer look, the density in a narrow under-dense region between two over-density regions is usually slightly biased high, for example, at $d\sim 3.5$ pMpc. This is mostly due to the thermal broadening of absorption lines.

To quantify the accuracy of the recovery process we show in the left panel of Figure \ref{fig:rspace_scatter} the $68\%, 95\%, 99.7\%$ contours of the scatter between the true gas density and the recovered gas density for all  unsaturated pixels in the range $0.5-4$ pMpc from the quasar. For most unsaturated pixels, the recovered gas density is very close the the true value. However, at $\Delta_g \sim 1$, we notice the contour has a ``bump". This over-prediction of the
gas density is mainly because the thermal broadening around overdensity peaks suppress the transmitted flux around them, biasing the density high. { In the right panel of Figure \ref{fig:rspace_scatter} we plot the same scatter plot but now with the true gas density smoothed by a Gaussian kernel with Doppler parameter $b\equiv\sqrt{2}\sigma=25 \rm km/s$. We can clearly see that taking the thermal broadening into consideration the scatter decreases, and there is no significant bias. We test Gaussian kernels with different Doppler parameters from $b=10\sim 30$ km/s and find that the recovered density agrees the best with the true density field smoothed by $b=25 \rm km/s$. (See also Section \ref{sec:T-rho}.)}

With this method, we can recover the density probability distribution function (PDF). By comparing density PDF at different distances from the quasar, we could potentially constrain how over-dense the quasar environment is. In the left panel of Figure \ref{fig:rspace_PDF} we show the PDFs for the true gas density (smoothed by $25 \rm km/s$) in different distance ranges. Since the halos we choose as the quasar hosts are the most massive ones in the simulation box, they trace a large-scale overdensity of $\sim 1$ pMpc in size. This is the reason that the purple line ($0.5-1.0$ pMpc) is significantly shifted to the right from other colored lines. On the right panel, we select three density ranges from the left panel and plot recovered density PDF from unsaturated pixels in dotted lines and the corresponding PDFs of the smoothed, true density field of these unsaturated pixels. The recovered density PDF traces the true PDF accurately, and the recovered PDF of $0.5-1.0$ pMpc shows significant overdensity compared with other PDFs at larger distances.

\subsection{Recovering gas density in redshift space}

\begin{figure*}
    \centering
    \includegraphics[width=\textwidth]{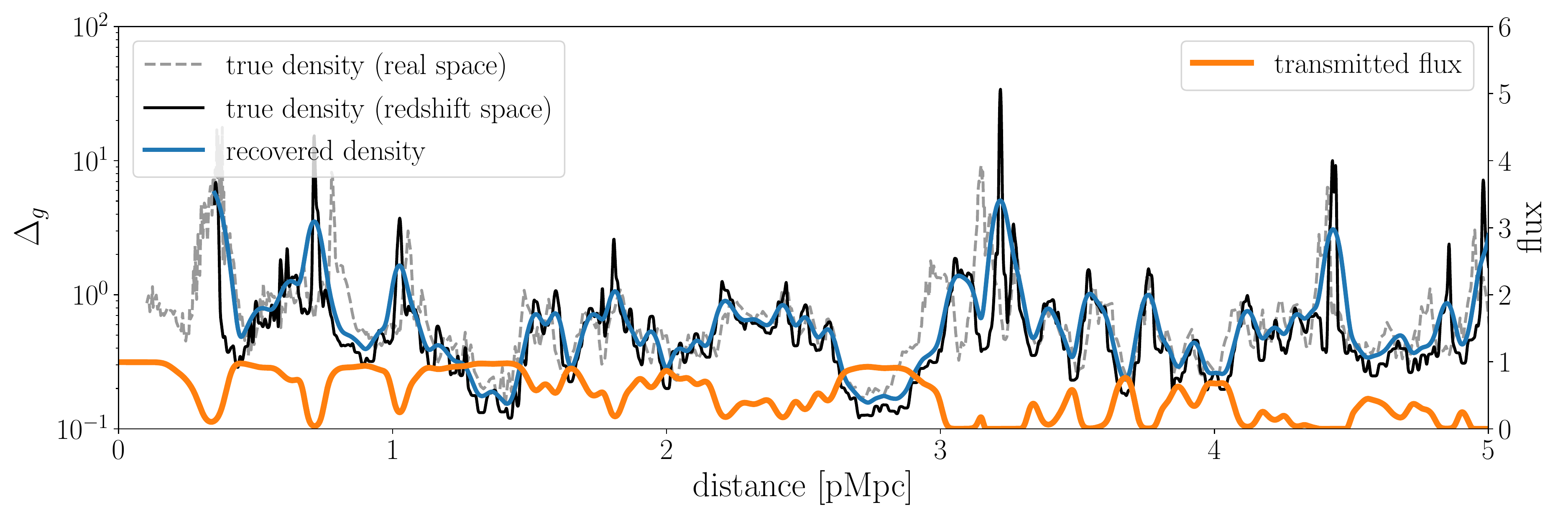}
    \caption{Same sightline as in Figure \ref{fig:novpec} but in a realistic case with gas peculiar velocities (solid lines). The black solid line is the true density in redshift space. The grey dashed line is in real space.}
    \label{fig:vpec}
\end{figure*}

Gas peculiar velocity is of the order of $100 ~\rm~ km/s$ at $z\sim6$, which can shift the absorption features in the redshift space by up to $\sim 0.2$ pMpc. This shift also introduces an error in the amplitude of the recovered gas density, because we only know the baseline optical depth at the corresponding real space location, not the actual redshift space location. 
This error scales as inverse distance. Therefore, for the recovered gas density relatively far from the quasar, say at $4$ pMpc, such an error is only $\sim 5\%$.

In Figure \ref{fig:vpec} we show the same sightline as in Figure \ref{fig:novpec}, but now with the full accounting for peculiar velocities. The gas density in redshift space is
\begin{equation} \label{Eq:tradDz}
    \Delta_{g,z}= \Delta_{g,r} \left| \frac{H dr}{dv} \right|=\Delta_{g,r} \left| \frac{H}{H+dv_{\rm pec}/dr} \right|.
\end{equation}
This density profile is shown with the solid dark black line in Figure \ref{fig:vpec} and the corresponding redshift-space density recovered using Equation \ref{Eq: Delta_g} is shown in solid blue.

Because the quasar host halo correlates with an overdensity on the scale of up to $\sim 1$ pMpc, which is collapsing as a whole, there is usually systematic gas inflow toward the quasar host halo on this scale. This can be seen from the shifts of the peaks at $d\approx0.8$ pMpc between the grey dashed and black solid lines in Figure \ref{fig:vpec}, which is a common feature for all the sightlines in our sample.

Most of density features are easily identifiable between the redshift space and the real space, despite shifts due to the peculiar velocity. Also, in the redshift space the density is usually lower in underdense region and higher in overdense region than in matching regions of the real space, as can be seen, for example, at $d\sim 3$ pMpc in Figure \ref{fig:vpec}. This is because the peculiar velocity usually has positive divergence ($dv_{\rm pec}/dr > 0$) in underdense regions and negative divergence ($dv_{\rm pec}/dr < 0$) in overdense regions.

Looking at the recovered density in blue, we find that it matches the true value generally well. However, in the large void at $d\approx 2.8$ pMpc, the blue line is higher than the dark black line. This is not solely caused by thermal broadening, which only impacts the gas around overdense peaks within $\sim 30$ pkpc. The degree of thermal effect in the big void at $d \sim 2.8$ pMpc can also be evaluated from Figure \ref{fig:novpec}, and is much smaller than that in Figure \ref{fig:vpec}. The major reason for the over-prediction of the density in such a large void is instead that Equation \ref{Eq: Delta_g} does not strictly recover the traditionally defined redshift density. Specifically, because the density of {\it neutral} hydrogen in redshift space is equal to the density of total hydrogen in redshift space times the neutral fraction in {\it real} space:
\begin{equation}\label{Eq:nHIz}
    n_{\rm HI, z}=x_{\rm HI} n_{\rm H, z}\propto \frac{\alpha n_{\rm H, r}}{\Gamma} n_{\rm H,z},
\end{equation}
where $n_{\rm H, z}$ and $n_{\rm H, r}$ represent the number density of total hydrogen in redshift space and real space, respectively. Therefore, the observable optical depth
\begin{equation}\label{Eq:tauHIz}
    \tau \propto n_{\rm HI, z} = x_{\rm HI} n_{\rm H, z}\propto  n_{\rm H,z} n_{\rm H,r}.
\end{equation}
Thus, by using Equation \ref{Eq: Delta_g} what we recover is in fact
\begin{equation}\label{Eq:tildeDelta}
 \sqrt{\Delta_r \Delta_z} \equiv  \Delta_{g,r} \sqrt{\left| \frac{Hdr}{dv} \right|}=
\Delta_{g,r} \sqrt{\left| \frac{H}{H+dv_{\rm pec}/dr} \right|},
\end{equation}
i.e.\ the geometric mean of the real space density and the redshift space density.

\begin{figure*}
    \centering
    \includegraphics[width=0.4\textwidth]{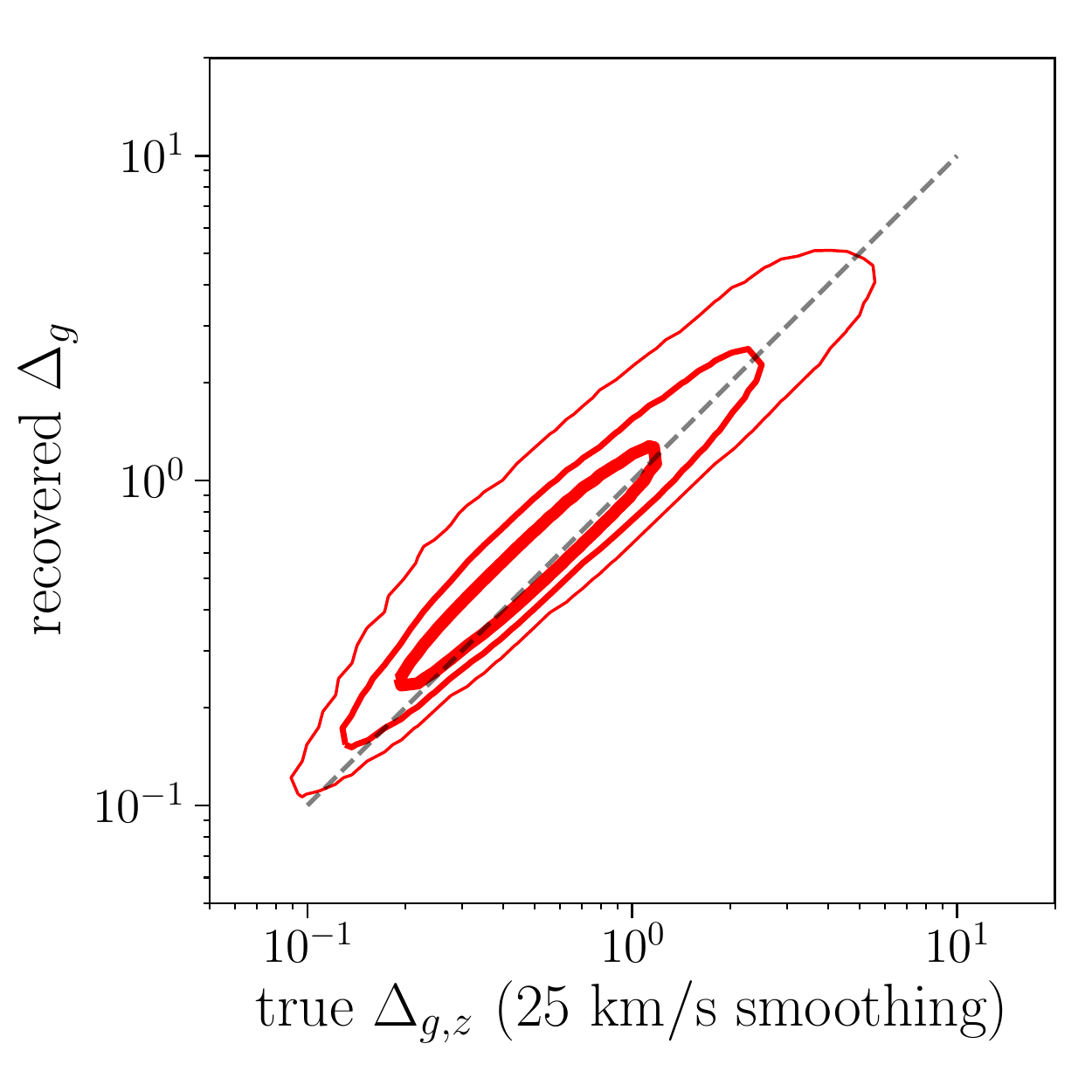}
    \includegraphics[width=0.4\textwidth]{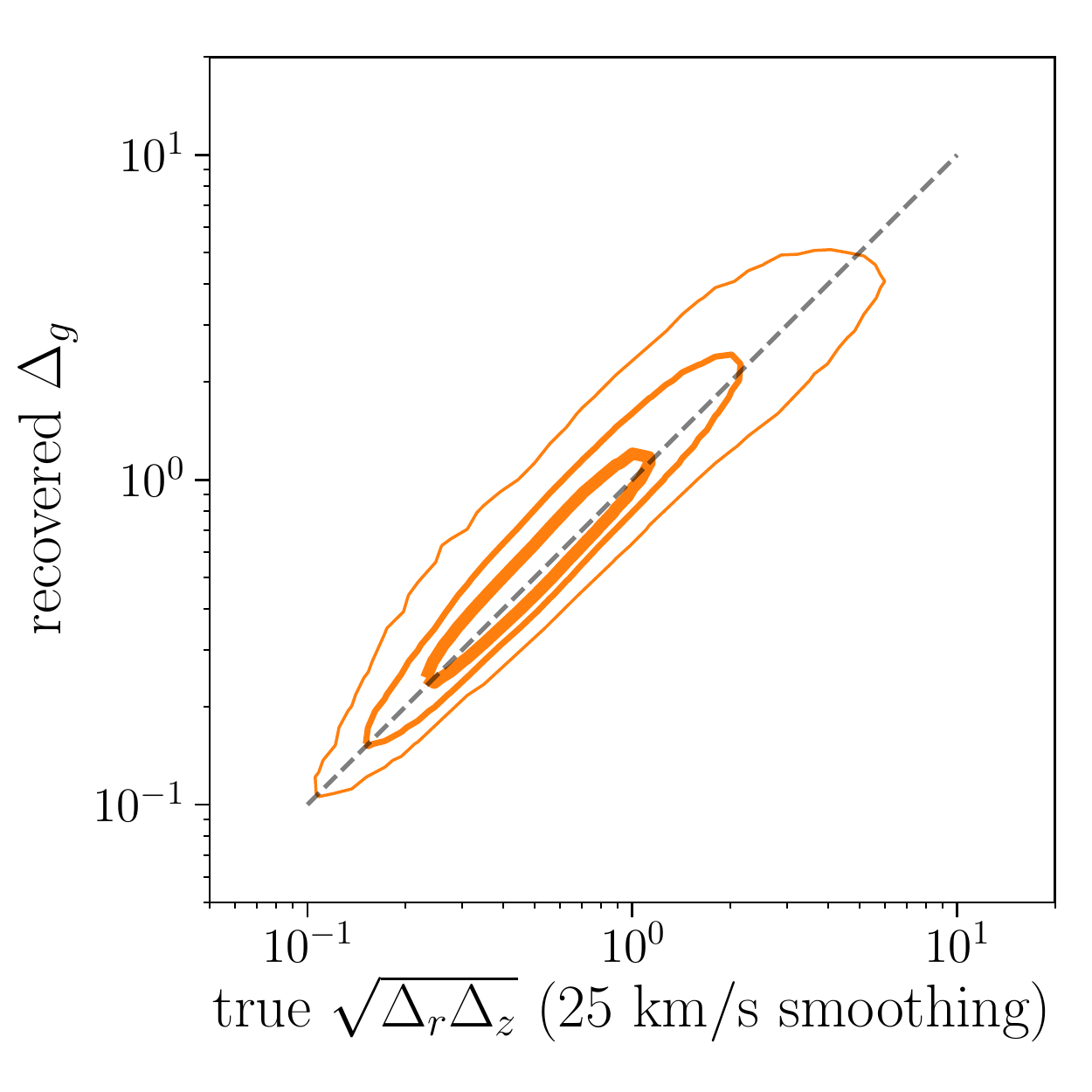}
    \caption{Scatter plot for the recovered gas density from synthetic spectra with peculiar velocity  versus the smoothed true gas density for all unsaturated pixels in the distance range of $0.5-4$ pMpc from all the 6001 quasars in our sample. The true density in the left panel is the usual redshift space density (Eq. \ref{Eq:tradDz}) while in the right panel it is the geometric mean of the real and redshift space densities as defined in Equation \ref{Eq:tildeDelta}.}
    \label{fig:zspace_contours}
\end{figure*}

\begin{figure*}
    \centering
    \includegraphics[width=0.45\textwidth]{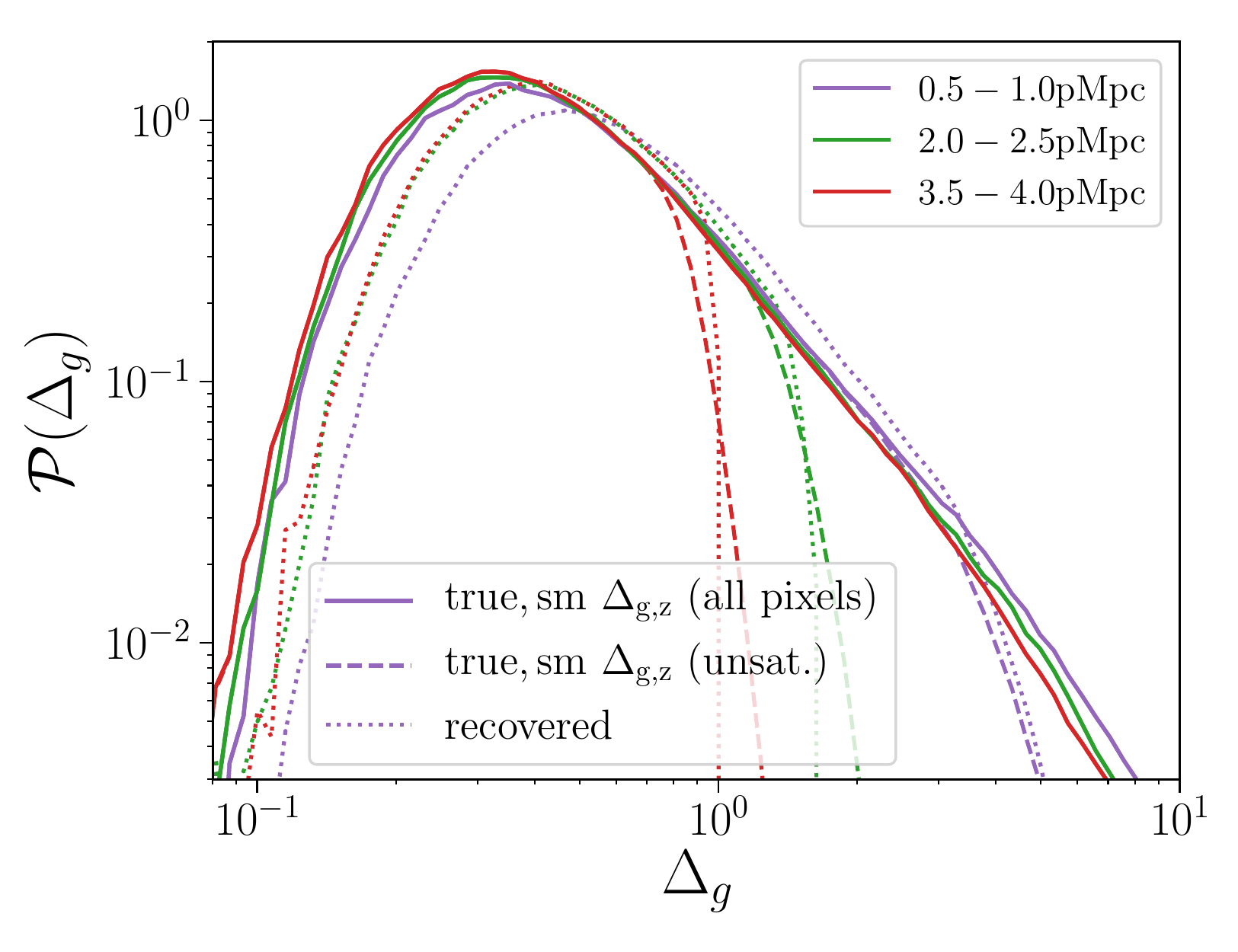}
    \includegraphics[width=0.45\textwidth]{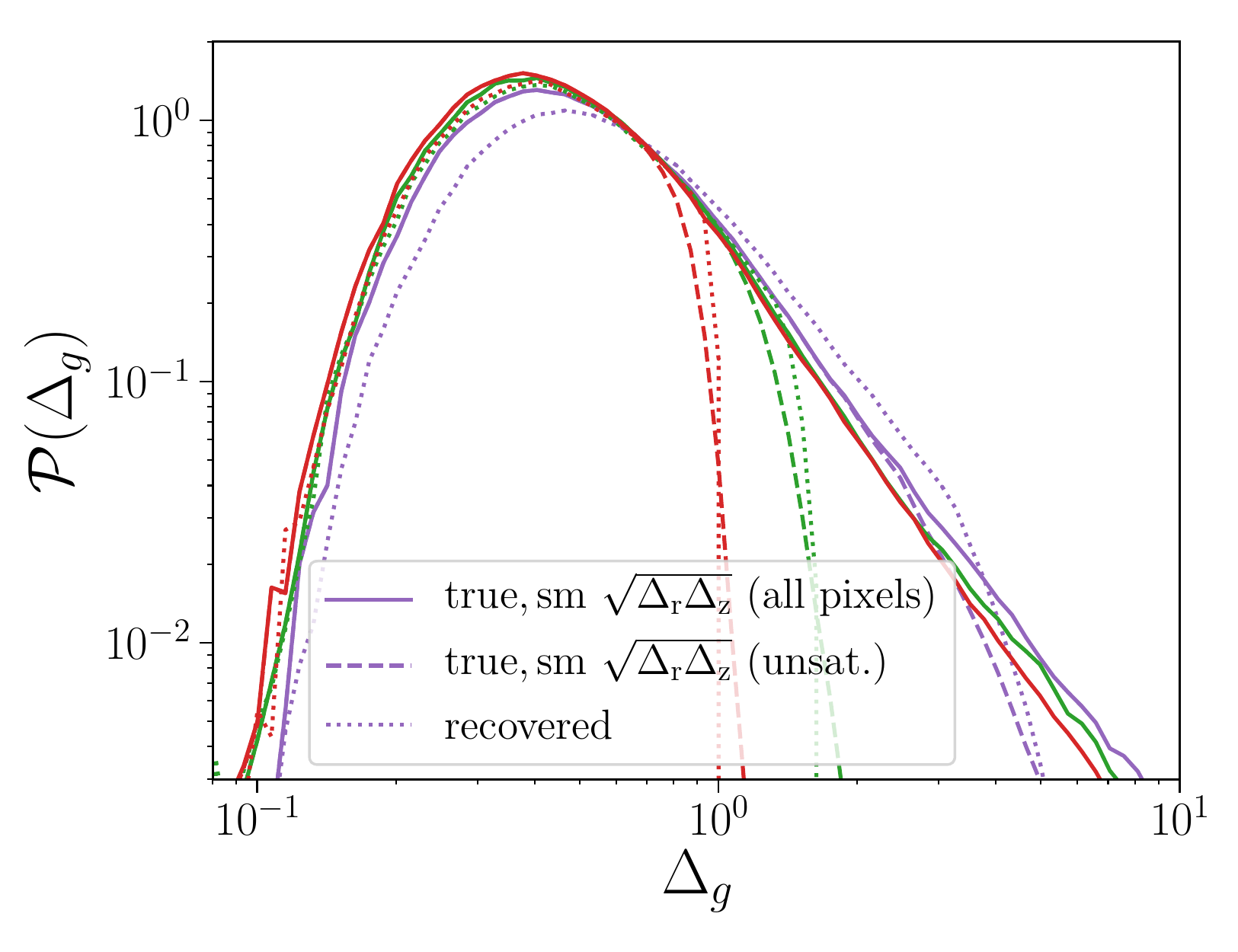}
    \caption{Same as the right panel of Figure \ref{fig:rspace_PDF}, but now synthetic spectra include peculiar velocity. Solid and dashed lines in the left panels are the smoothed redshift space density (as defined in Eq. \ref{Eq:tradDz}) and in the right right panels are the geometric mean of the real and redshift space density as defined in Equation \ref{Eq:tildeDelta}.}
    \label{fig:zspace_PDF}
\end{figure*}

We quantify the accuracy of the recovered redshift space density with the $68\%, 95\%, 99.7\%$ confidence contours in Figure \ref{fig:zspace_contours}. The y-axis of both panels in the figure are recovered gas density contrast with Equation \ref{Eq: Delta_g}. In the left panel, the x-axis is the smoothed (with a $25 ~\rm~ km/s$ Gaussian kernel) true traditionally defined redshift space density. Compared with Figure \ref{fig:rspace_scatter}, the scatter is larger and the low density end is biased high. As shown previously, the bias at low densities is because Equation \ref{Eq: Delta_g} is not recovering the traditionally defined redshift density, but the one defined as Equation \ref{Eq:tildeDelta}. In the right panel, we plot the smoothed gemetric mean of the  redshift and real space densities as defined in Equation \ref{Eq:tildeDelta}, and the bias at low densities essentially disappears.

Similarly to the right panel of Figure \ref{fig:rspace_PDF}, in the left panel of Figure \ref{fig:zspace_PDF} we show with solid lines the true smoothed PDF of the redshift space density at three different distances from the quasar. Compared to the real-space density in Figure \ref{fig:rspace_PDF}, the PDFs of the redshift space density (solid lines) are wider because low density regions tend to have coherent expansion and high density density often contract, shifting the redshift space density further way from the value of the mean density of the universe.
The recovered PDFs (dotted lines) are systematically shifted towards higher densities because of the bias explained previously. In the right panel of Figure \ref{fig:zspace_PDF} we show the PDFs of the geometric mean of the real and redshift space density as defined in Equation \ref{Eq:tildeDelta}. As expected, the recovered density PDFs agree much better with the true underlying PDFs of $\sqrt{\Delta_r \Delta_z}$. The only detectable deviation is in the first distance bin $0.5-1.0$ pMpc, which is due to the systematic inflow of gas toward the quasar within $\approx 1$ pMpc from the quasar host halo. We briefly note that this bias may be larger for more massive halos than that in our simulations, which are all below $2\times 10^{12} ~\rm~\Msun$, but we are unable to test this directly in our simulation due to the limited box size.

\section{Discussion} \label{sec:discussion}

In reality, many factors may impact the accuracy of density recovery.
For example, it is not clear how well we can determine the continuum of the quasar; errors in the continuum introduce errors in the measured transmitted flux. Also, HeII photo-heating from the quasar increases the IGM temperature and, hence, affect the neutral fraction, introducing yet another error. In this section, we focus on these two systematic errors and briefly discuss some other potential systematics. As we will see, different systematics impact recovered density PDF differently, which may offer a unique way to constrain quasar properties.

\subsection{Uncertainty in Continuum Modeling}

\begin{figure*}
    \centering
    \includegraphics[width=\textwidth]{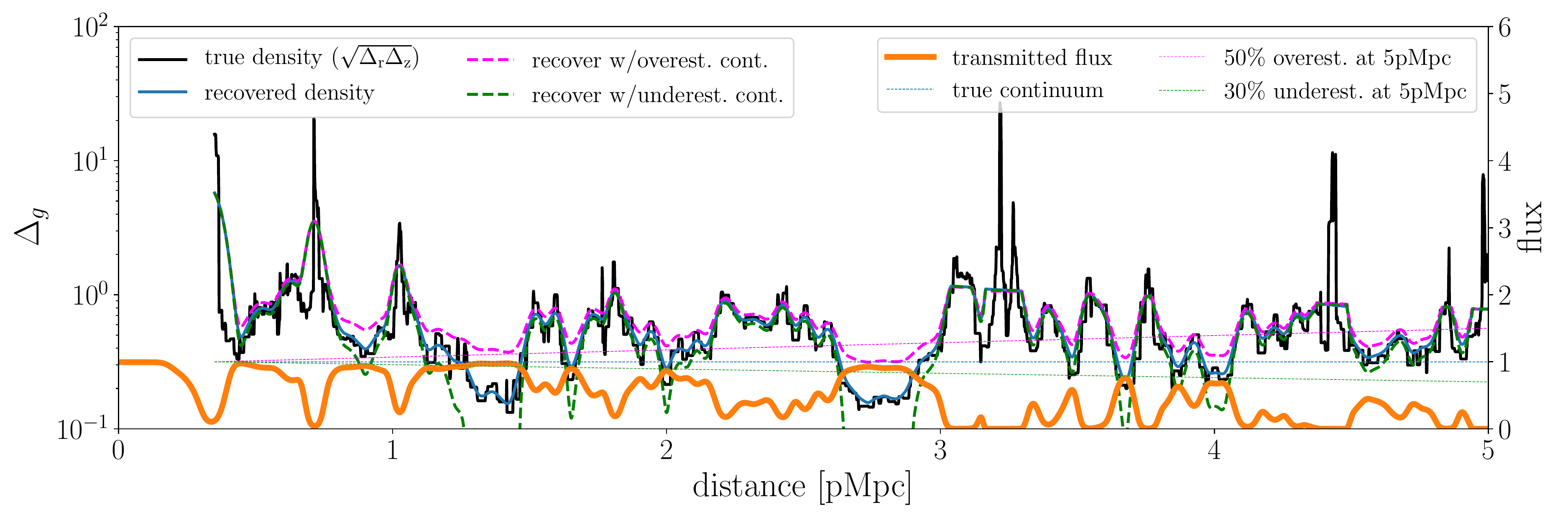}
    \includegraphics[width=0.33\textwidth]{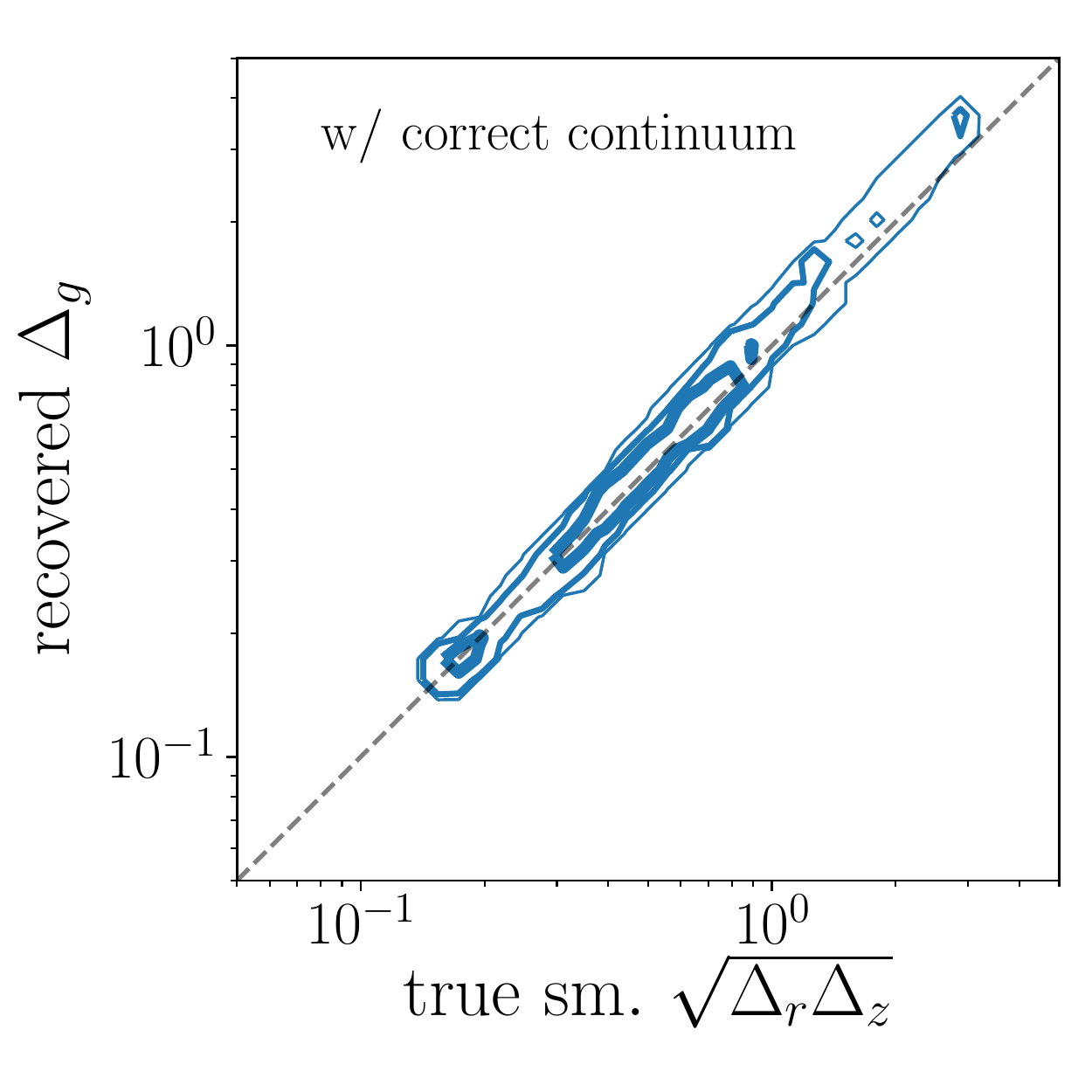}
    \includegraphics[width=0.33\textwidth]{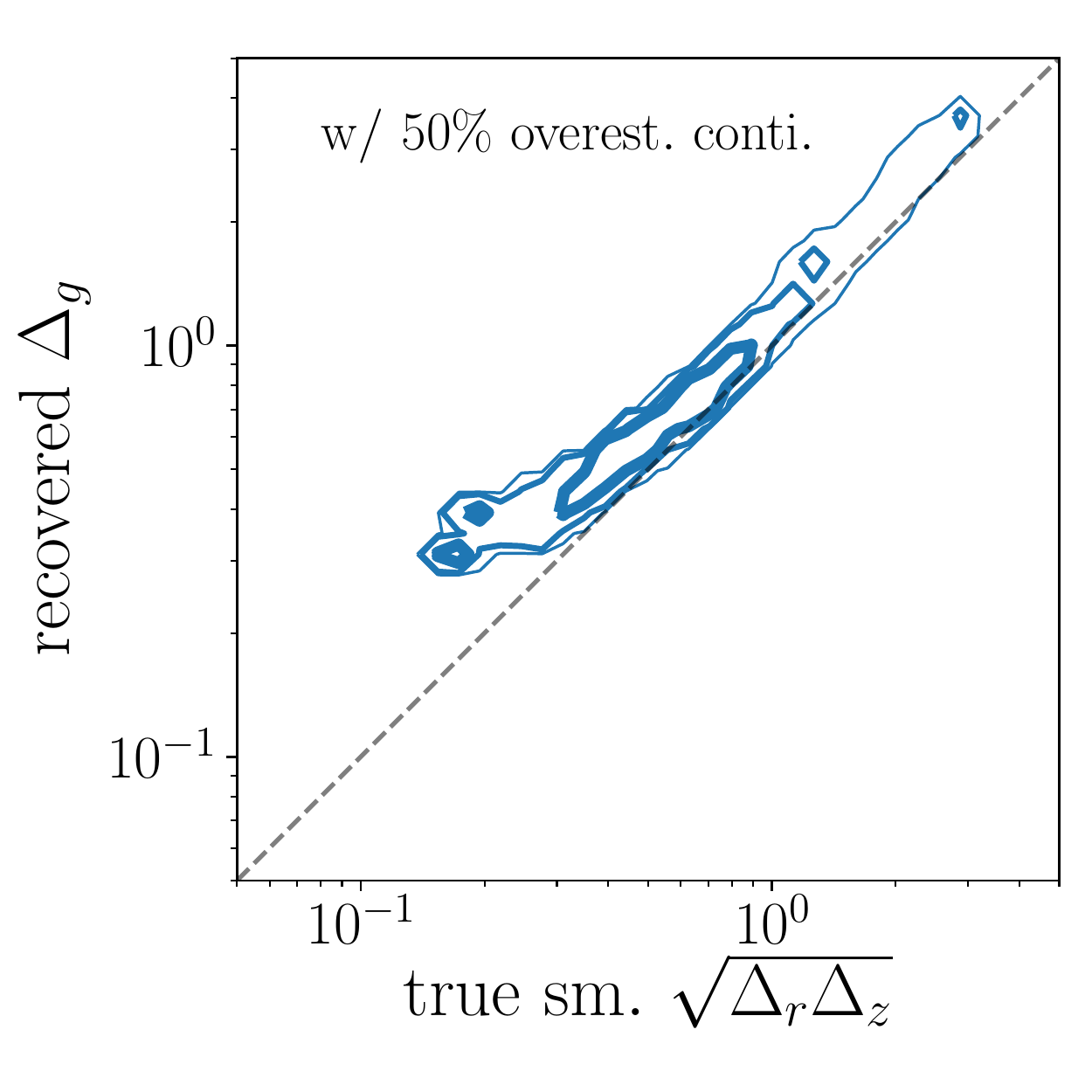}
    \includegraphics[width=0.33\textwidth]{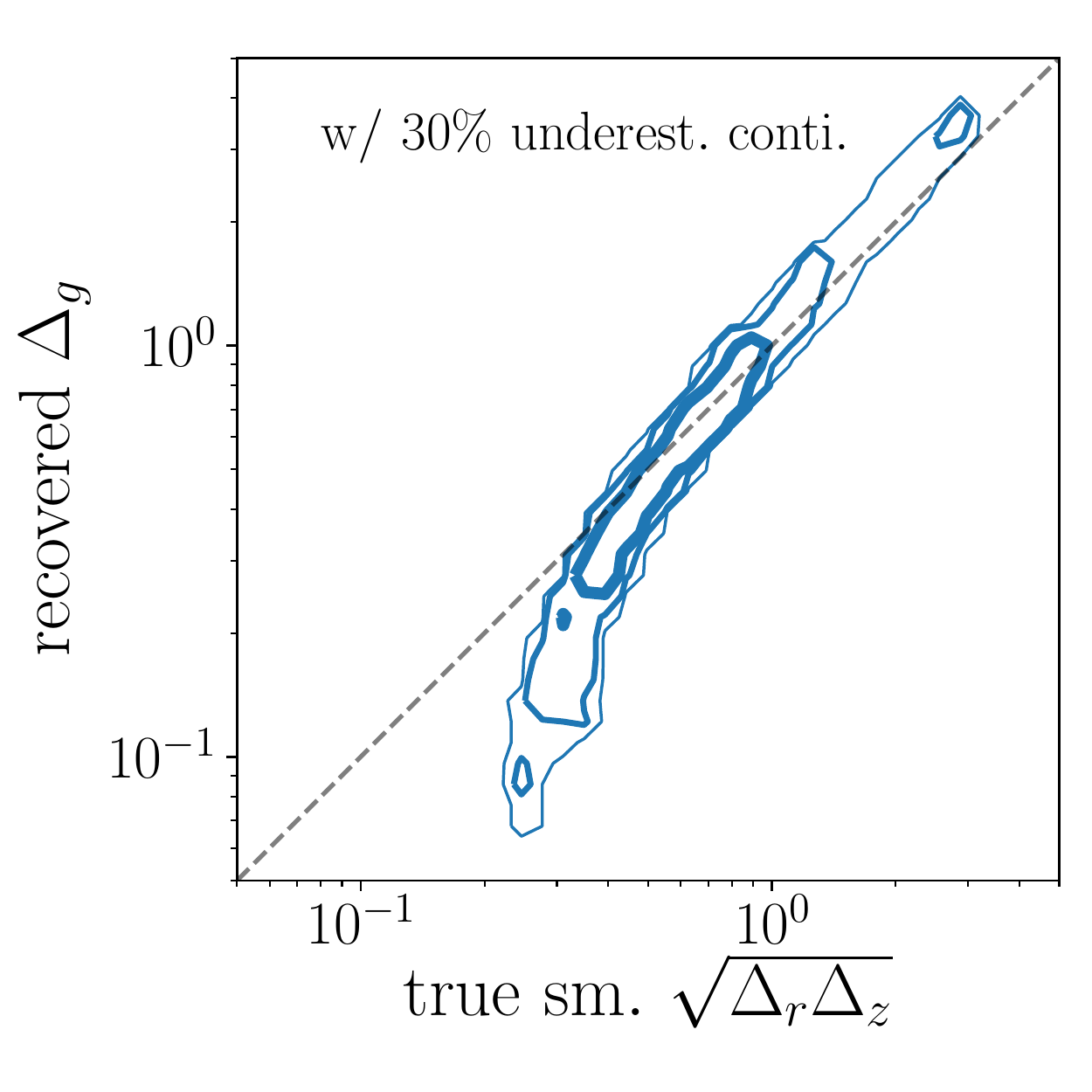}

    \caption{Upper panel: the same sightline as in Figure \ref{fig:novpec}. The black line is the true geometric mean of the real and redshift space densities $\sqrt{\Delta_r \Delta_z}$ and the blue line is the recovered density using the correct continuum. The blue, magenta and green lines are the correct, overestimated and underestimated continua. The magenta and green dashed lines are the recovered density profiles using the corresponding over- and under-estimated continua. Lower panels: for all unsaturated pixels between $0.5\sim 4$ pMpc from the upper panel, the $68\%, 95\%, 99.7\%$ contours of true $\sqrt{\Delta_r \Delta_z}$ smoothed by a $b=25 ~\rm~ km/s$ Gaussian kernel vs.\ recovered density with the correct, over- and under-estimated continua shown in the upper panel. }
    \label{fig:errcon}
\end{figure*}

In practice, the quasar continuum is often modelled using the spectra redward of the Ly$\alpha$ line. However, either a traditional PCA or a machine learning method may introduce systematic errors.  In order to mimic this uncertainty and quantify how it may impact the density recovery, we introduce a systematic error in quasar continuum that linearly depends on the distance, as shown by the straight magenta and green lines in Figure \ref{fig:errcon}.  This error is normalized so that at $5$ pMpc from the quasar the continuum is overestimated by $50\%$ (magenta) or underestimated by $30\%$ (green), and is correct close to the quasar at $0.35$ pMpc. Note that this is a very conservative choice, as in real observations the continuum errors are usually lower than this \citep[see][]{davies2018,bosman2020}. In Figure \ref{fig:errcon} we show the same sightline as in Figure \ref{fig:vpec}, with the black line showing the true geometric mean of the real and redshift space densities as defined in Equation \ref{Eq:tildeDelta}. Using the true continuum, we recover this value very well, as is shown by the solid blue line and also by the left panel below. With the over- or under- estimated continua, the density recovery suffers. The error is especially severe in the underdense regions, where the relative error in the optical depth is large. In contrast, gas with $\Delta_g>1$ has relatively larger absorption and thus is less sensitive to the continuum errors.

\begin{figure}
    \centering
    \includegraphics[width=0.45\textwidth]{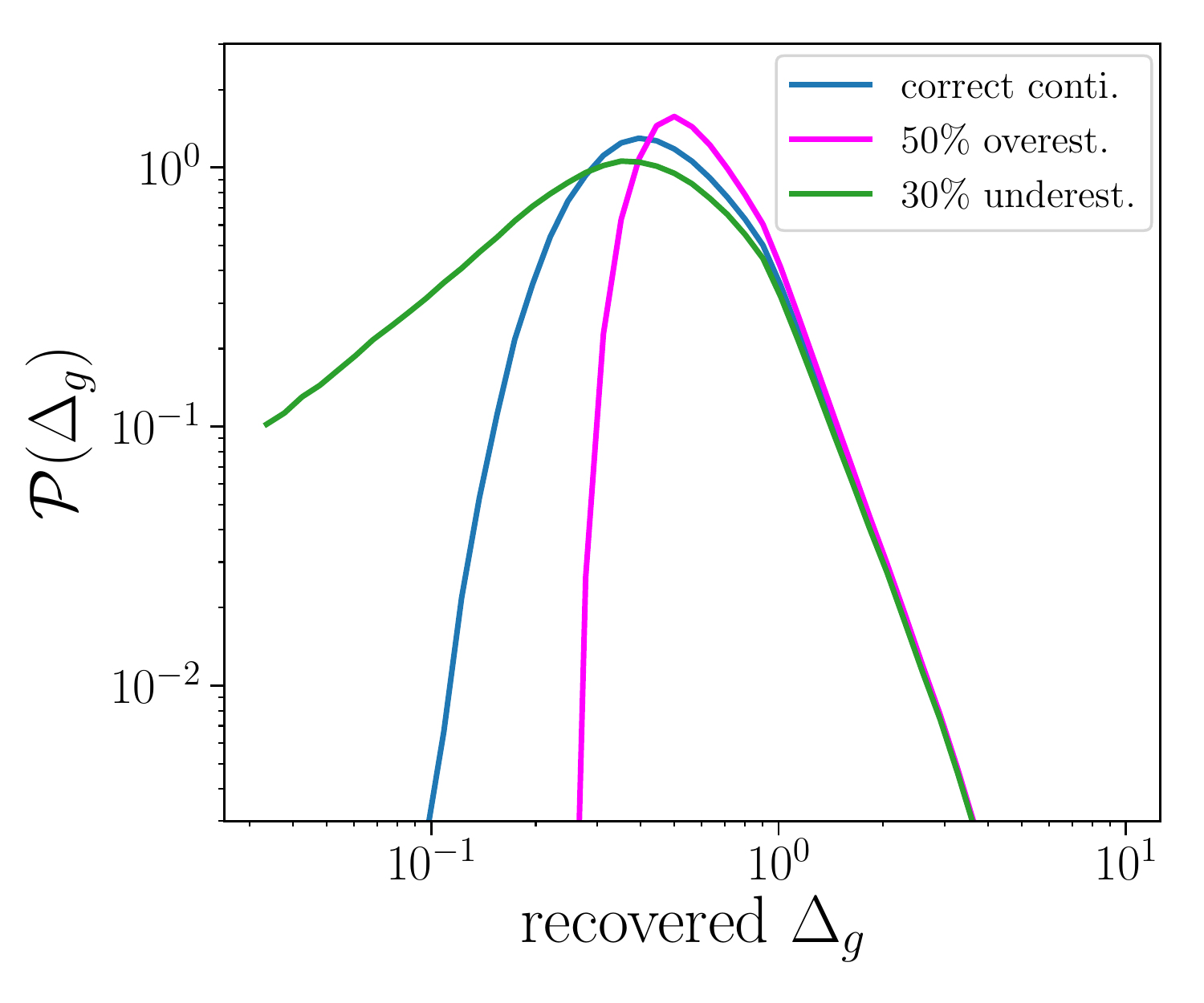}
    \caption{The density PDFs of recovered density for all unsaturated pixels within $0.5-4$ pMpc for all the  sightlines in our sample. The blue, magenta and green colors correspond to the density recovered using the correct, overestimated and underestimated continuums in Figure \ref{fig:errcon}.}
    \label{fig:contiSysPDF}
\end{figure}

We use the over- and under-estimated continua to recover all $6001$ sightlines in our sample, and In Figure \ref{fig:contiSysPDF} we show the recovered density PDFs for all unsaturated pixels within $0.5-4$ pMpc. The blue is recovered using the true quasar continuum, while the magenta and the green are using the over- and under- estimated continuum described above.
Using the correct continuum, the lowest density pixels should be larger than about $0.1$. With the underestimated continuum, however, the density PDF has an extended low density tail, which can even have negative densities when transmission spikes exceed the local continuum and appear as ``emission'' rather than absorption in the spectrum. On the other hand, with the overestimated continuum, the lower end ($\Delta_g<1$) of the PDF is artificially pushed toward higher densities. At densities above the cosmic mean the three PDFs are not  distinguishable though.

\subsection{Systematics in Baseline Modeling due to Quasar Ages} \label{sec: age} 

The density recovery formula (Equation \ref{Eq: Delta_g}) is based on Equation  \ref{eq:alphaT}, where we approximate $T$ as being uniform after reionization. This is likely to be a good assumption in most places, because right after rapid ionization the IGM temperature mainly depends on the shape of the ionizing spectrum but not its amplitude. Currently observations \citep[e.g.][]{onoue2017, matsuka2018} suggest that galaxies, which commonly have soft spectra, are the dominating reionization sources. Because galaxies do not have enough photons with energy above $4$ Rydberg, most of helium at $z\sim6$ should be singly ionized. In our CROC simulation, PopII stars are the main reionization sources, and the single ionized helium fraction at $z=6$ is $x_{\rm HeII}\approx 0.9$.

However, quasars themselves have copious amounts of HeII ionizing photons that can doubly ionize helium in their immediate surroundings and raise the IGM temperature even more. Such temperature increment reduces the recombination rate of hydrogen and slightly impacts its the ionization fraction, affecting the transmitted flux within quasar proximity zones \citep{bolton2007, padmanabhan2014}. Depending on the age of the quasar, a quasar HeIII region may well be confined to its proximity zone but not be small enough as to be unimportant. The non-trivial temperature structure induced by the HeIII region in turn induces deviations in the baseline model and introduces a systematic error. 

\begin{figure*}
    \centering
    \includegraphics[width=0.24\textwidth]{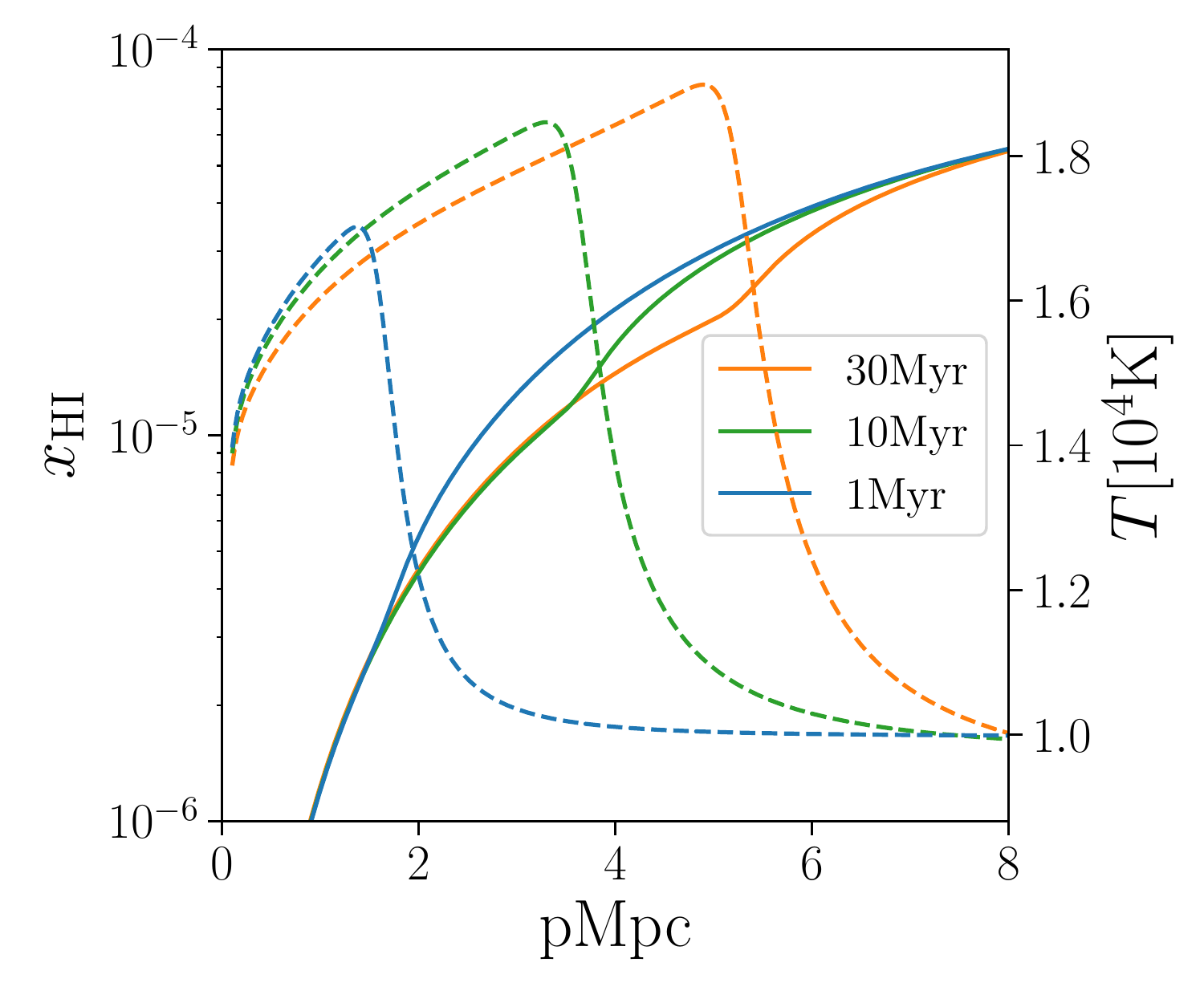}
    \includegraphics[width=0.24\textwidth]{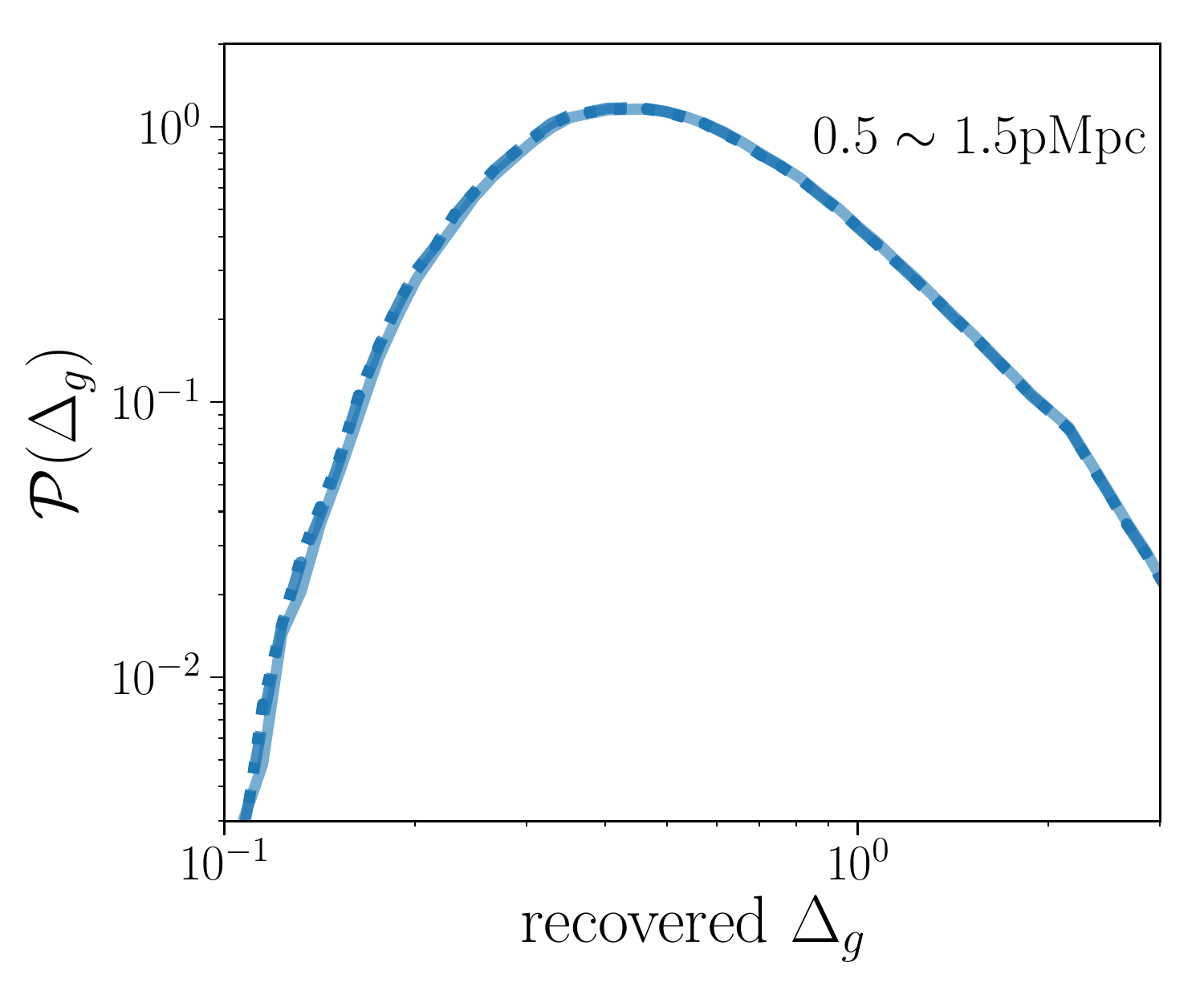}
    \includegraphics[width=0.24\textwidth]{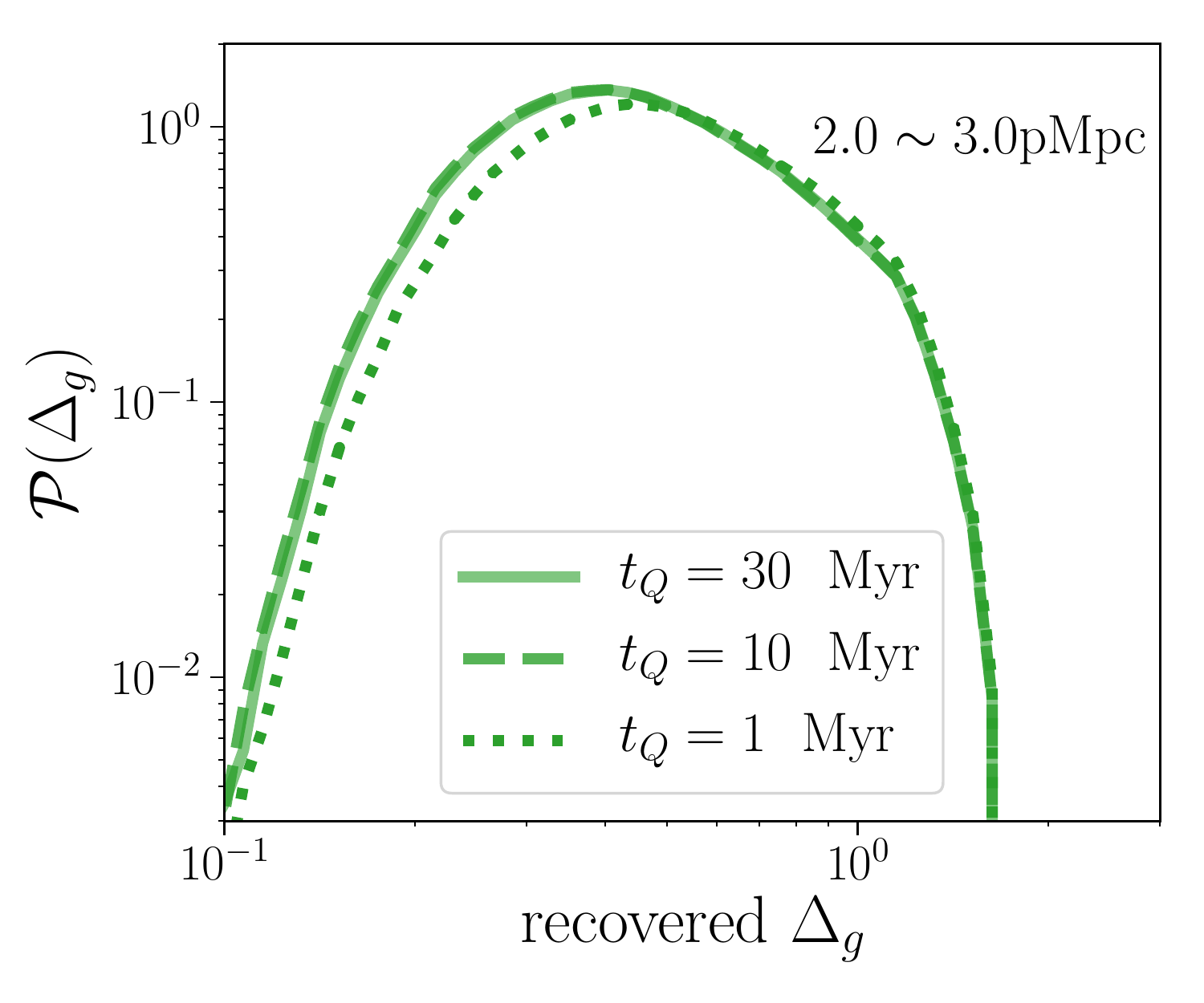}
    \includegraphics[width=0.24\textwidth]{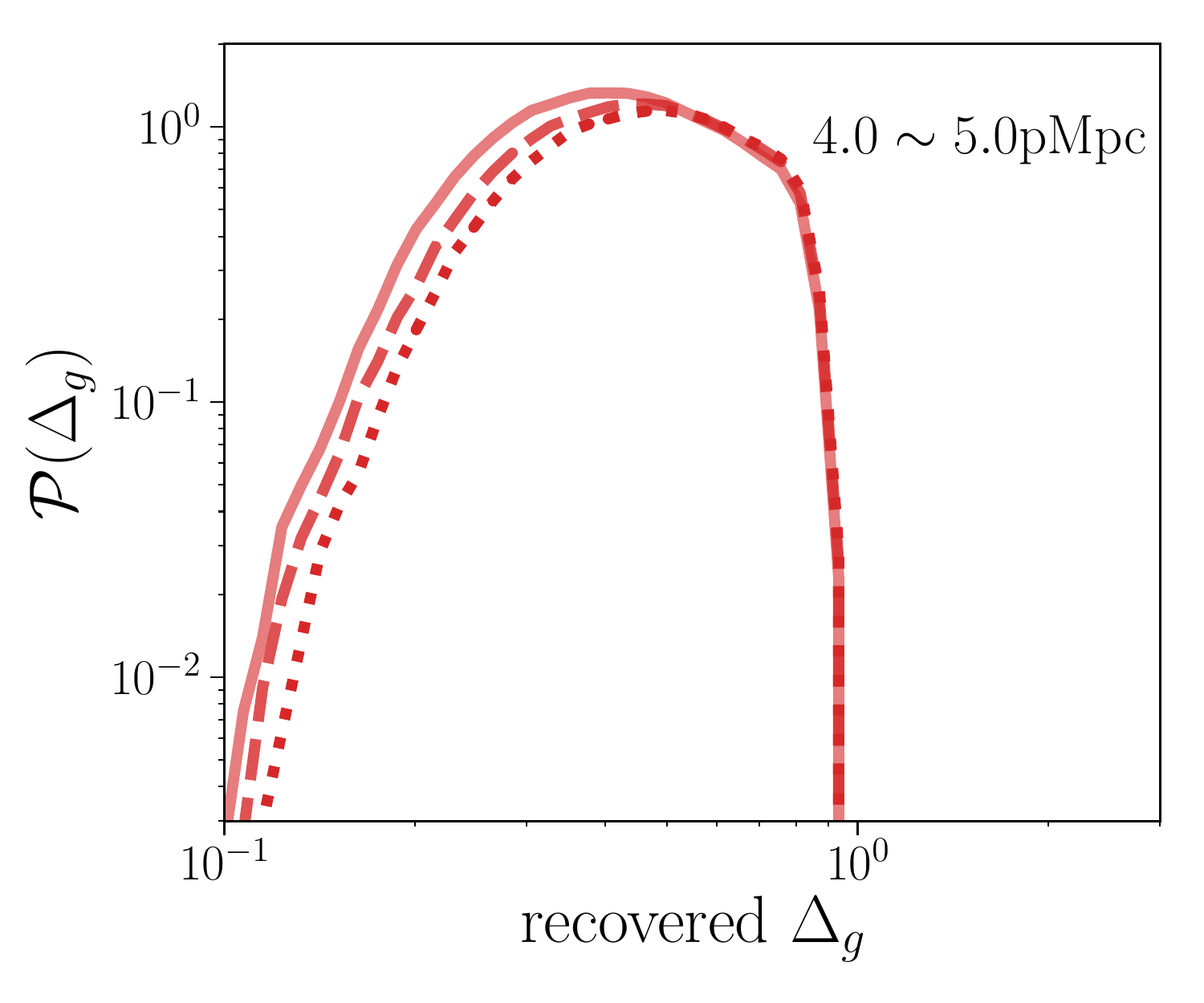}
    \caption{Left: neutral fraction (solid lines) and gas temperature (dashed lines) at quasar age $t_Q=30$ Myr, $10$ Myr, and $1$ Myr, assuming a uniform IGM with the cosmic mean density and the initial conditions of the cosmic mean at $z=6.11$ in the simulation. Right three panels: recovered density PDFs using the fiducial baseline of $t_Q=30$ Myr applied to the same sample of sightlines with different quasar ages. The solid, dashed, and dotted lines represent $t_Q=30$ Myr, $10$ Myr, and $1$ Myr, respectively. From left to right the density PDFs are shown in radial shells at distances $0.5\sim1.5$ pMpc, $2\sim3$ pMpc, and $4\sim5$ pMpc, respectively. }
    \label{fig:tQpdf}
\end{figure*}

To examine how quasar age impacts the density recovery procedure, we apply the recovery method with the fiducial baseline for $t_Q=30$ Myr to the same sample of sightlines but with the assumed quasar ages of $t_Q=1$ Myr and $t_Q=10$ Myr. 
In the leftmost panel of Figure \ref{fig:tQpdf} we show the neutral hydrogen fraction (solid lines) and the gas temperature (dashed lines) in a uniform density universe with different quasar ages. This is to show an average position of the HeII I-front of a $\dot{N}=1\times10^{57}~\rm~s^{-1}$ quasar with a spectral index of $-1.5$. 
As the quasar age increases from $1$ Myr to $10$ Myr and to $30$ Myr, the HeII front position moves from $\sim 2$ pMpc to $\sim 4$ pMpc and to $\sim 6$ pMpc.
The temperature inside the HeII I-front is $\sim 5000 {\rm K}$ higher than immediately outside it. Therefore, using the fiducial baseline of $t_Q=30$ Myr we expect to see a systematic error in recovered density fields at $2\sim 6$ pMpc for a young quasar of $t_Q=1$ Myr and at $4\sim 6$ pMpc for $t_Q=10$ Myr.

To quantify this systematic error, in the other panels of Figure \ref{fig:tQpdf} we show the recovered density PDF at different distance bins.
For the distance bin of $0.5\sim 1.5$, which the HeII I-front has swept within $1$ Myr, there is no significant dependence on the quasar age. However, for the distance bin of $2\sim 3$ pMpc the recovered density for a young quasar of $t_Q=1$ Myr is systematically higher than that of $t_Q=30$ Myr, while the PDFs of the recovered densities are similar for $t_Q=10$ Myr and $t_Q=30$ Myr, after the HeII front has passed. For the farthest distance bin of $4\sim 5$ pMpc the recovered density PDF of both $t_Q=1$ and $10$ Myr are significantly larger than that for $t_Q=30$ Myr, but the difference is larger for $t_Q=1$ Myr. Note that the HeII I-front has a certain width. The peak of the temperature profile often corresponds to the position where the ionization fraction $x_{\rm HeII}\approx 0.01$. The temperature drops significantly and approaches the initial value at the point where $ x_{\rm HeII}\approx0.8$. The width of this transition region from $x_{\rm HeII}\approx 0.01$ to $x_{\rm HeII}\approx 0.8$   is $\sim 1$ pMpc. At $t_Q=10$ Myr, even though the HeII I-front has not passed beyond $4$ pMpc  (the point where $x_{\rm HeII}=0.5$ is at $3.7$ pMpc), the temperature has already risen somewhat at distances up to 5 pMpc. On the other hand, when the quasar is only $t_Q=1$ Myr old, the HeII I-front is so close to the quasar that the temperature at $4\sim 5$ pMpc is not affected at all. This explains why the $t_Q=10$ Myr PDF has a smaller shift at the  $4\sim 5$ pMpc bin than the $t_Q=1$ Myr one.

\subsection{Sample Variance, Spectral Noise, and Resolution}

\begin{figure}
    \centering
    \includegraphics[width=0.45\textwidth]{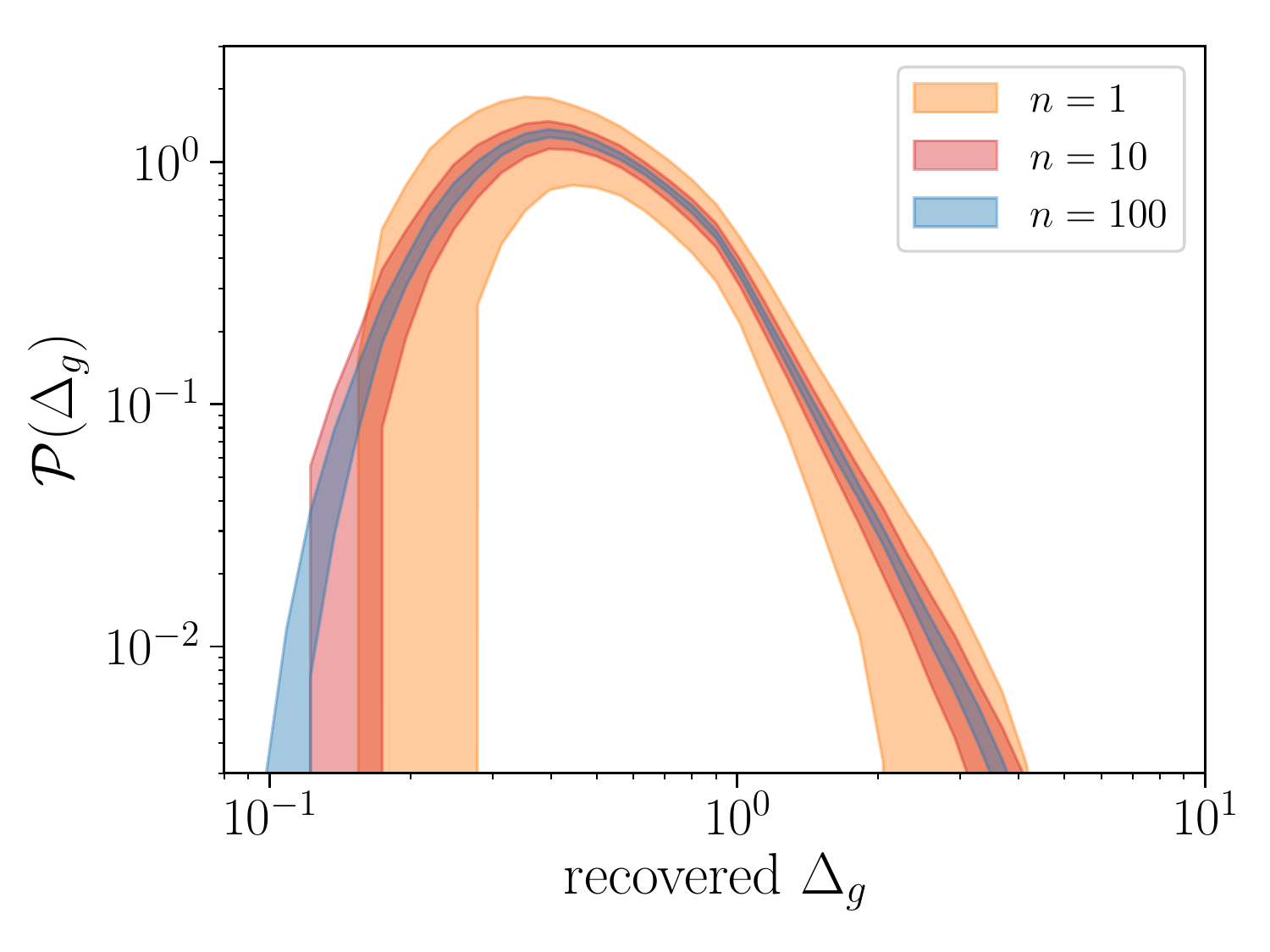}
    \caption{ Density PDF recovered with $1, 10$ and $100$ sight lines, respectively. The envelopes show the $68\%$ uncertainty.}
    \label{fig:sample_variance}
\end{figure}

{ In practice, the number of $z>6$ quasar spectra is not unlimited. According to the observed quasar luminosity function at $z=6$ \citep{willott2010,jiang2016,onoue2017}, the number density of quasars with $M_{1450}<-26$ is $\approx 10^{-9} \rm Mpc^{-3}$. Considering the comoving volume between $z=5.5 \sim 6.5$ $V_{z6}\approx390 \rm Gpc^3$, there are potentially a few hundred bright quasars that can be used to recover the density field at $z\sim 6$. In this subsection, we investigate the sample variance for realistic quasar samples.

We calculate the sample variance of the recovered density PDF between $0.5-4$ pMpc by randomly drawing sets of  $n$ ($n=1,10,100$) sight lines. 
For each $n$, we draw at least $6000/n$ sets and plot the $68\%$ envelope of the recovered density PDF in Figure \ref{fig:sample_variance}. 
With only one sight line, the uncertainty is large. 
However, with the increasing number of sight lines the uncertainty shrinks as $1/\sqrt{n}$, as expected for Poisson errors.
When $n$ increases from $10$ to $100$, the uncertainty reduces from $20\%$ to $6\%$. Therefore, it is hopeful that we can recover the density PDF at $z\sim 6$ with statistical error $\sim 5 \%$ with the existing observational technology. 
Obviously, the statistical error will become even smaller when 30-meter class telescopes come online.

Noise and resolution of the spectra, as always, introduce extra uncertainties into the results. From Equation (\ref{Eq: Delta_g}), in order to achieve $10\%$ precision in $\Delta_g$, the signal-to-noise must be above $5$ inside the proximity zone. 
As for the spectral resolution, currently the best resolution spectra of $z\sim 6$ quasar is $R\gtrsim 10000$, which is able to reach the theoretical limit ($b= 25$ km/s) of the smallest scale density fluctuation we can recover. If there is extra broadening from instrument, it will erase some small scale structure but is not expected to bias the recovered PDF. As a test, we smooth synthetic spectra additionally with a Gaussian filter of width $b=75$ km/s and compare the recovered PDF with the PDF of the underlying density field smoothed with the same scale. We find that the agreement is still very good. For significantly larger instrumental broadening, however, the true cosmological density fluctuations on the corresponding scale are too small to be recovered. Thus, using low resolution spectra ($R<4000$) may not provide informative results. 
}

\subsection{ Temperature-Density Relation} \label{sec:T-rho}
\begin{figure}
    \centering
    \includegraphics[width=0.45\textwidth]{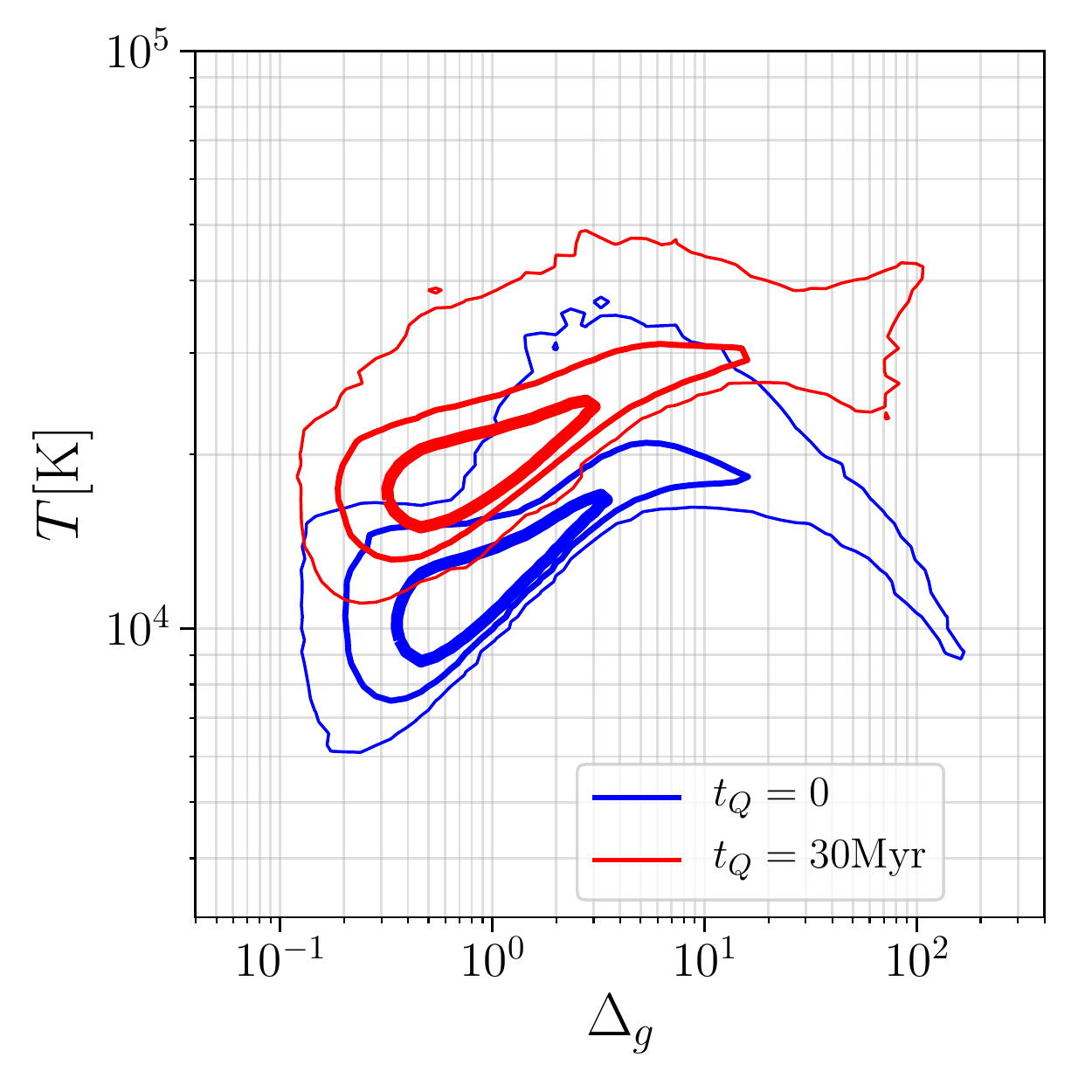}
    \caption{ The temperature-density relation before the quasar turns on (blue contours) and after the quasar has been shining for $30$ Myr (red contours). Contours show the $68\%, 95\%$ and $99.7\%$ of pixels within $0.5-4$ pMpc.}
    \label{fig:T-rho}
\end{figure}

{
Transformation from Equation \ref{Eq: tau} to \ref{Eq: Delta_g} assumes that the recombination rate is constant for the gas at different densities. However, even though at the end of cosmic reionization the $T-\Delta_g$ relation is expected to be flat (and which will be even flatter after the gas is further ionized by the quasar), it is still not exactly a constant. In Figure \ref{fig:T-rho}, we show the $T-\Delta_g$ relation for pixels within $0.5-4$ pMpc from the quasars for
the $6001$ sight lines we use in this work. Without additional ionization from the quasar, the power-law slope of the $T-\Delta_g$ relation is around $0.25$, and the slope is reduced to $0.2$ after the quasar has been shining for $30$ Myr.

Considering a temperature-density relation of $T-\Delta^{\gamma-1}$ and $\alpha(T)\propto T^{-0.7}$, we can come up with a generalized density recovery formula:

\begin{equation}\label{Eq: general}
\Delta_g = \left( {\frac{{\tau}_{\rm Ly \alpha}}{\bar{\tau}_{\rm Ly \alpha}}} \right)^{\frac{1}{2-0.7(\gamma-1)}}.
\end{equation}

Using the slope measured above, a possibly more accurate recovery formula is 

\begin{equation}\label{Eq: precise}
\Delta_g = \left( {\frac{{\tau}_{\rm Ly \alpha}}{\bar{\tau}_{\rm Ly \alpha}}} \right)^{\frac{1}{2-0.7(\gamma-1)}}=
\left( {\frac{{\tau}_{\rm Ly \alpha}}{\bar{\tau}_{\rm Ly \alpha}}} \right)^{1/1.86}.
\end{equation}

We test how the recovery result changes with Equation (\ref{Eq: precise}). Here we use the spectra without peculiar velocity and compare the recovered density with the smoothed real space density. In the left panel, we plot the recovered density with the new formula against the true density smoothed by $25$ km/s, similar to the right panel of Figure \ref{fig:rspace_scatter}. We find that there is a slight bias, especially at the low density end. However, the smoothing scale that we initially choose, $b=25$ km/s, corresponds to the gas temperature of $3.8\times 10^4$ K. This is higher than the  actual temperature of the bulk of the gas. A more physically motivated smoothing scale should be $b\approx 15$ km/s, corresponding to $T\approx 1.8\times 10^4$ K, the temperature of gas whose density is recoverable ($\Delta_g\lesssim 1$; non-saturated Ly$\alpha$ absorption). In the right panel, we plot the recovered density against the true density smoothed by $b=15$ km/s, and we find that the agreement improves slightly at the lower end. 
This exercise indicates that the reason that the recovered density with Equation \ref{Eq: Delta_g} agrees the best with the true density smoothed by the larger scale of $b=25$ km/s is because there is a degeneracy between $\gamma$ in the generalized density recovery formula Equation \ref{Eq: general} and the best fit smoothing scale.
}

\begin{figure*}
    \centering
    \includegraphics[width=0.45\textwidth]{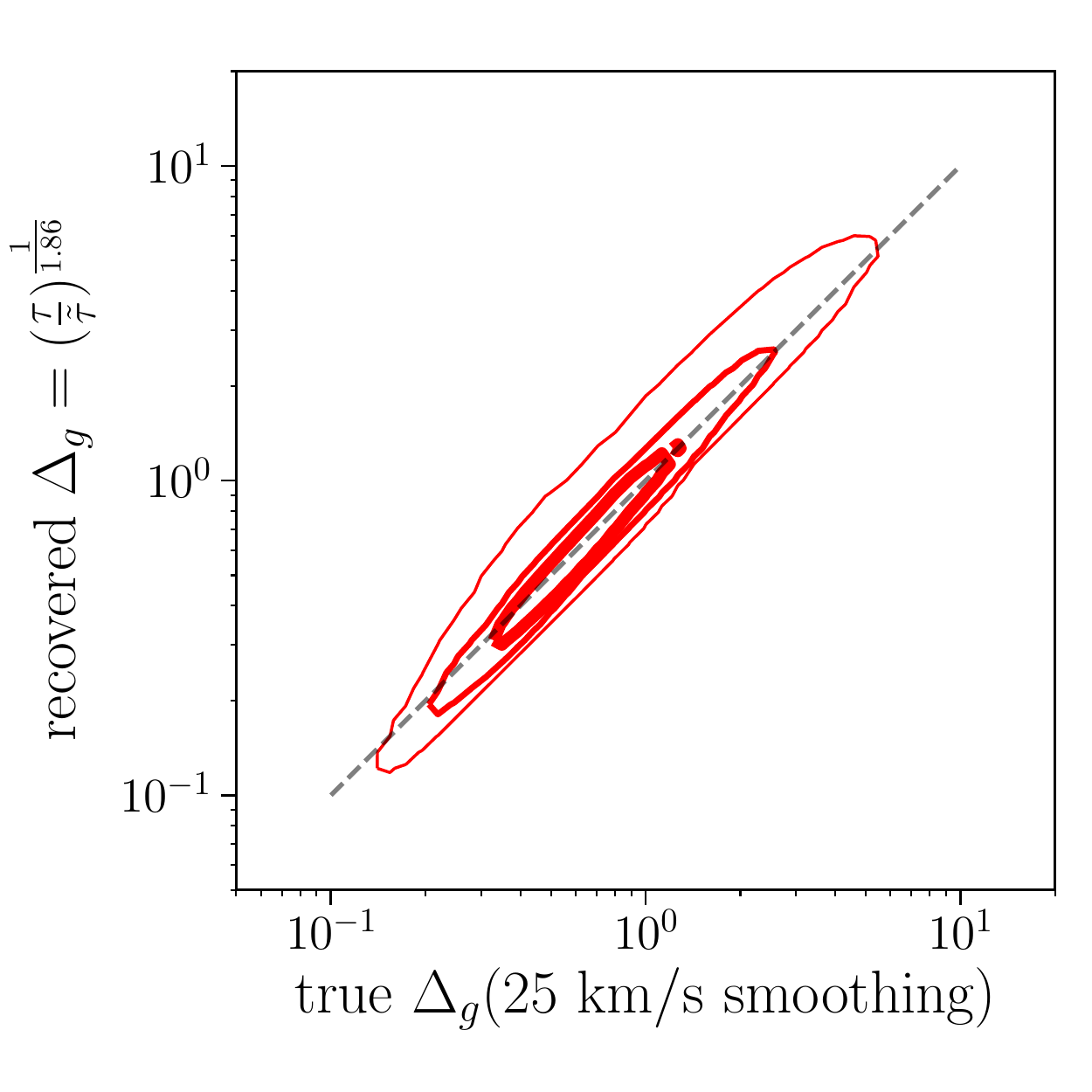}
    \includegraphics[width=0.45\textwidth]{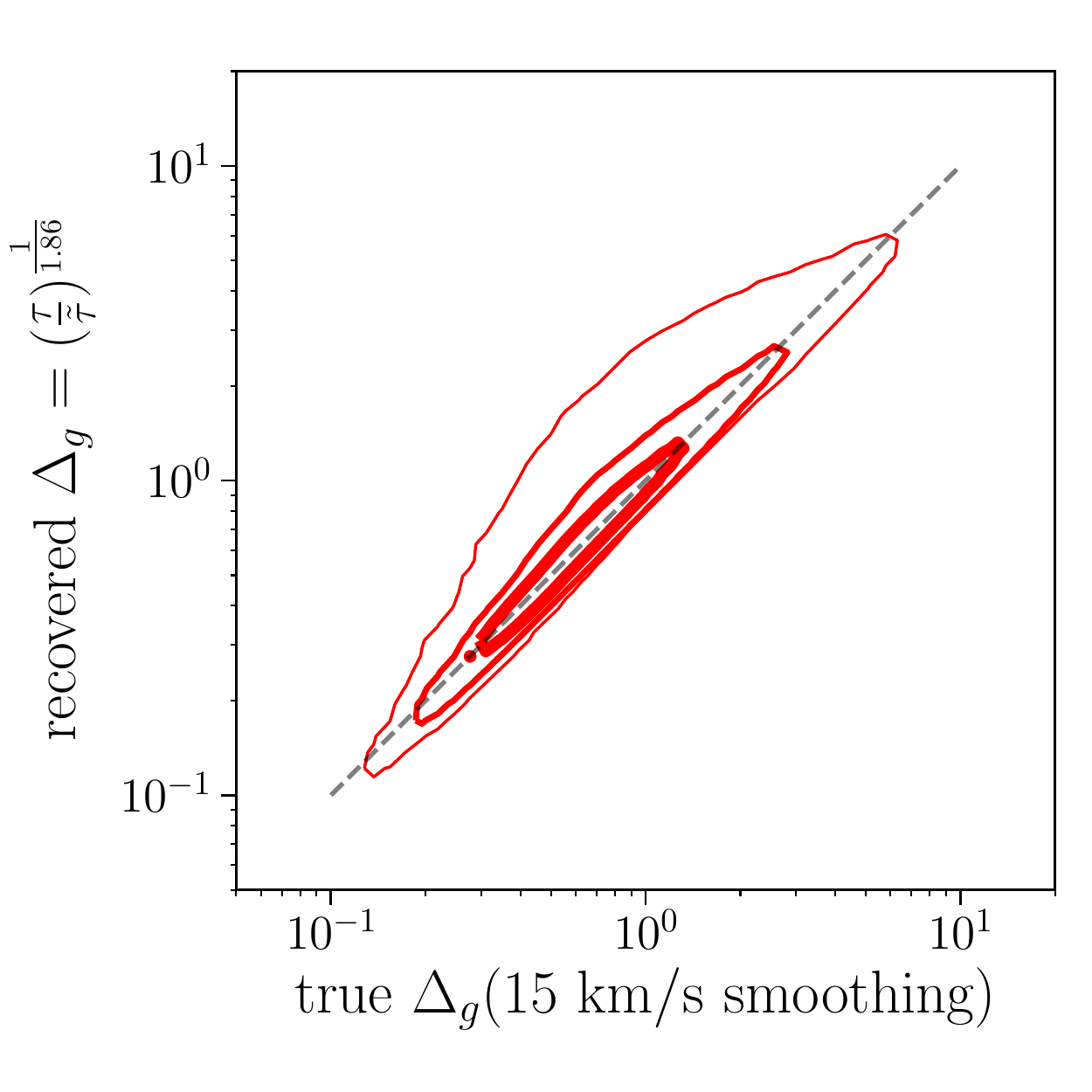}
    \caption{ Left: Density recovered using the adjusted formula Equation \ref{Eq: general} versus true density smoothed by a Gaussian kernel of $b=25$ km/s. Right: the same as left but versus true density smoothed by a Gaussian kernel of $b=15$ km/s.}
    \label{fig:general_rec}
\end{figure*}

\subsection{Caveats}
The density recovery procedure is based on the assumption that we know exactly what the ionizing flux of the quasar is and that this flux remains constant. 
In real life the magnitude of the quasar is measured at the red side of the Ly$\alpha$ line, often at the rest-frame of $1450 $ \AA, because there is no significant absorption or emission there.
However, the exact spectrum blueward of $912$ \AA\ is not directly observable, due to the heavy absorption from residual neutral hydrogen at $z\sim 5$. 
Although at $z<4$ the absorption from the IGM is limited and the continuum can be estimated relatively well \citep[e.g.,][]{kirkman2005}, at $z=4\sim 5$ there still exists large uncertainties.
Also, it is not straight-forward to extrapolate the continuum across $\sim 912$ \AA\ because often the is a break in the slope. \citet{lusso2015} suggests that the uncertainty in the quasar photoionizing rate can be as high as $\sim 20\%$ due to the uncertainty in the spectral slope. Since the optical depth is inversely proportional to the photoionization rate (Equation \ref{Eq: tau}), the recovered gas density contrast should have an uncertainty of $\sim 10\%$ due to uncertainly in the quasar ionizing luminosity; that uncertainly, however, does not depend on distance. { The uncertainty in the spectral slope also has a secondary effect on photoheating. We test that by running a sight line with the uniform density and a much harder spectrum with a spectral index of $-0.5$ and the same $\dot{N}=1\times 10^{57} \rm s^{-1}$. Compared with  the fiducial spectral of index $-1.5$, we find an increase in the temperature of $18\%$ at $d=0.5$ pMpc and of $27\%$ at $d=4$ pMpc. This translates into a systematic $\sim 10\%$ error in $\alpha(T)/\alpha(\tilde{T})$ (Equation \ref{Eq: tau}) and propagates to a $\sim 5\%$ error in the recovered density, which depends weakly on distance. If we use the average spectral index of $-1.7$ reported in \citet{lusso2015}, compared with the fiducial spectral index of $-1.5$, the change in the temperature is $\sim 3\%$ and the corresponding change in the recovered density is $\sim 1\%$.}

Quasar variability in ionizing flux can break down the ionization equilibrium assumption. According to Figure \ref{fig:attenuationFrac}, if the quasar is stable for $t_Q>1$ Myr, this method should be safe for a quasar with $M_{1450}<-26.67$ within $\sim 4$ pMpc. However, if $t_Q$ is smaller than $\sim 1$ Myr, the recovered density would be biased. Unlike the scenario of a stable quasar with the incorrect ionizing flux discussed in last paragraph, the bias due to quasar variability should increase with distance, because the ionization timescale is larger at larger distances. If a quasar is switching on and off quickly with episodic lifetimes $\lesssim 0.1$ Myr, even the size of the traditionally measured proximity zone ($R_{\rm PZ, obs}$) would be impacted \citep{davies2020}. Therefore, the method here should not apply to quasars with sufficiently small proximity zones. For a quasar with $R_{\rm PZ, obs}$ around average or above, the method should be relatively safe for at least the inner half of the proximity zone, because the very fact that a quasar has the proximity zone of an average size implies that the IGM around the edge of the proximity zone should be close to the ionization equilibrium. That being said, how exactly this bias depends on the quasar duty cycle is worth further exploration.

\begin{figure}
    \centering
    \includegraphics[width=0.45\textwidth]{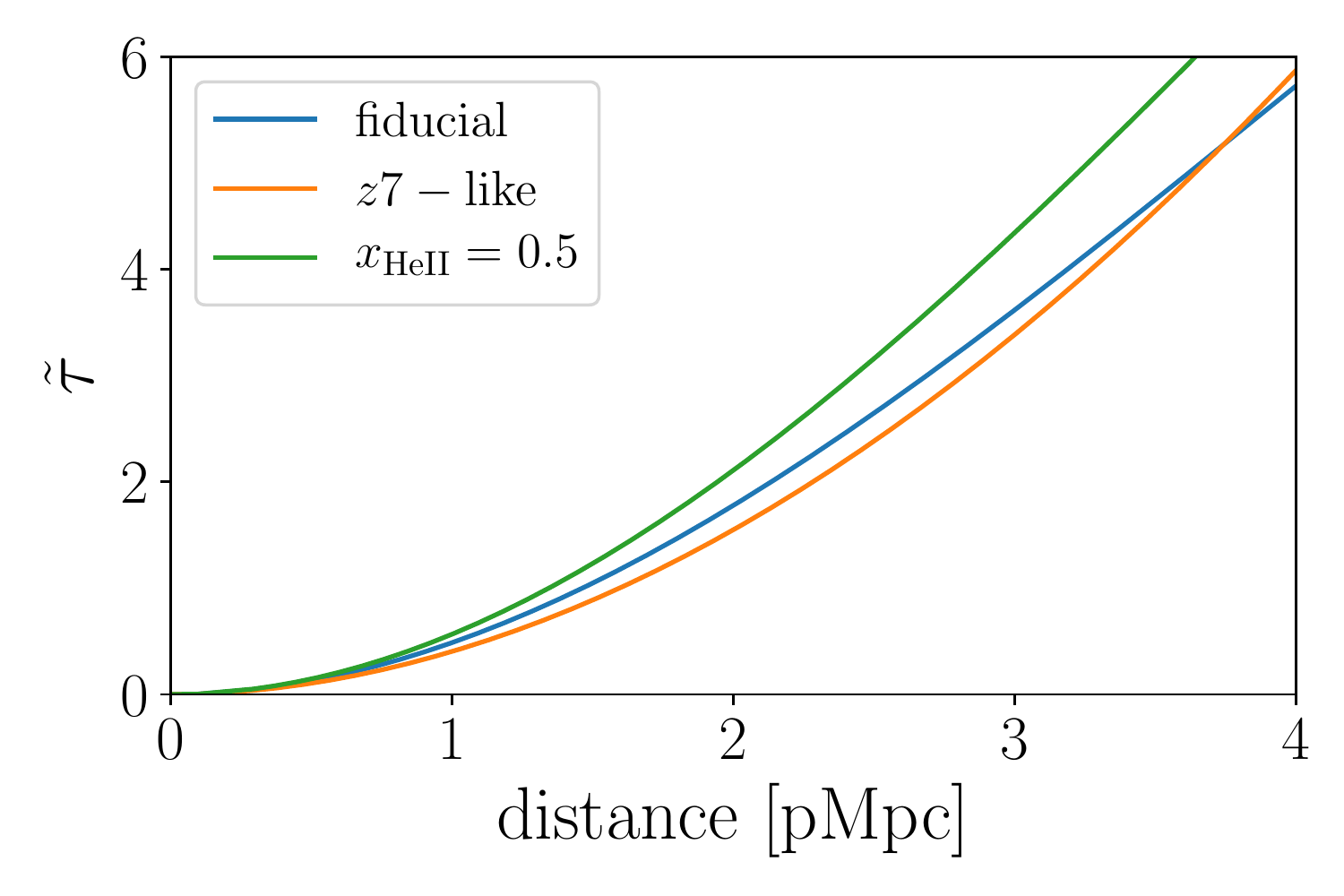}
    \caption{Baseline models with different initial IGM properties before the quasar turns on. The blue line is the fiducial baseline using the mean value of IGM at $z=6$. The orange line is using $z7$-like initial value. The green line is the same as the fiducial one except $x_{\rm HeII}=0.5$. All of them are with $t_Q=30$ Myr.}
    \label{fig:REIhist}
\end{figure}

One other factor that may affect the recovered density is the initial condition we use when calculating the baseline. With different ionization fraction or temperature prior to the quasar cycle, the IGM temperature (and thus the transmitted flux) after quasar switches on may change too.  Here we only estimate the amplitude of this effect by running several baselines with different physically motivated values. We explore two sets of initial IGM condition:
\{$x_{\rm HI}=0.001, x_{\rm HeI}=0.002, x_{\rm HeII}=0.98, T=15000~\rm~K$\}, which resembles the IGM in the simulation at an earlier redshift $z\sim 7$, right after reionization, and \{$x_{\rm HI}=10^{-4}, x_{\rm HeI}=10^{-4}, x_{\rm HeII}=0.5, T=10000~\rm~K$\}, which is the same as the fiducial model except half of the helium is doubly ionized, mimicking a reionization with much harder spectra. Note that by adopting these initial values, we also modify the background ionization rates\footnote{We calculate the background ionization rate, which is fixed constant, by assuming the ionization equilibrium using the initial ionization fraction and temperature.}: The $z7$-like baseline is calculated with the lower background photoionization rates and the $x_{\rm HeII}=0.5$ one has a higher background HeII photoionization rate. In Figure \ref{fig:REIhist} we show the baselines of these two variations compared with the fiducial one. The differences we see in $\tilde{\tau}$ mainly come from different HII recombination rates $\alpha(T)$. With higher initial $x_{\rm HI}$, $x_{\rm HeII}$ and temperature, the $z7$-like baseline has higher final temperature, a lower recombination rate, and thus lower $\tilde{\tau}$ well inside the proximity zone. On the contrary, the $x_{\rm HeII}=0.5$ baseline has less photoheating from HeII, lower temperature, and therefor higher $\tilde{\tau}$. The differences in baseline $\tilde{\tau}$ from these different reionization history models are $\sim 10\%$, and would propagate to the recovered gas density.

\subsection{Potential usage}

The reconstructed density field is a potentially powerful tool to study the quasar environment and quasar physics itself. The most challenging task is perhaps to obtain a robust quasar continuum. However, once we obtain a confident continuum, by using the statistics of the reconstructed density field we can constrain the quasar properties, since different properties impact the reconstructed density differently, as is shown above. We summarize how quasar properties impact the recovered density PDF in Table \ref{tab:potential}. The uncertainties in quasar ionizing flux and quasar age bias the PDF in different ways. The former impacts the whole recovered density field independently of distance. On the other hand, the latter results in a bias jump across the HeII I-front. More specifically, if the quasar is younger than expected, the recovered density would be biased high $\it outside$ the HeII I-front. Besides, examining the PDF also helps us constrain the quasar environment, since if the quasar resides in a massive halo (which correlates with the density field on $\sim$ pMpc scale), the density $\it inside$ the clustering scale is higher than the mean. This again has a different signature. What exact statistics we should use to extract those information is worth exploring in the future.

\begin{table}[]
    \centering
    \begin{tabular}{|c|c|}
    \hline
        Properties   &  Signiture \\
    \hline
       quasar ionizing flux  &  bias independent of distance \\
     \hline
       quasar age ($\gtrsim 1$ Myr)  & bias jump across HeII I-front \\
      \hline
      quasar host halo mass & denser PDF inside the clustering length \\
      \hline
    \end{tabular}
    \caption{}
    \label{tab:potential}
\end{table}

\section{Conclusion}

We study how to recover density field within the quasar proximity zone using synthetic spectra. The spectra are obtained by post-processing a CROC simulation. We have found that inside the proximity zone, the majority of quasars display a completely transparent IGM, thus the density can be recovered with a simple formula. Using this method, the recovered density agree very well with the true geometric mean of the real and redshift space densities $\sqrt{\Delta_r\Delta_z}$ with a scatter of $20\%$. We can recover density around the cosmic mean for a quasar of $\dot{N} \sim 1\times 10^{57} ~\rm~ s^{-1}$ (or $M_{1450}\sim -26.67$ assuming a $-1.5$ spectral index above $1450$ \AA) and put lower limits on the density of saturated features. 

We examine various systematic errors. Uncertainties in determining the quasar continuum can greatly impact the recovered density in underdense regions.  Another bias comes from  photoheating of HeII: if the position of the HeII I-front differs from the baseline model, the recovered density is biased across the HeII I-front. We also emphasize that this method is based on the assumption that the quasar has a stable ionizing flux and has been shining for at least $\sim 1$ Myr. If the quasar is flickering with short lifetime and displays an exceptionally small proximity zone, the IGM may not be in the ionization equilibrium and our method breaks down.

Our density recovery method has many potential uses. For example, by comparing the density PDF at different distances, we can potentially constrain the density environment around the quasar. The recovered density field in the quasar proximity zone can thus offer us a unique tool for understanding where and how the first quasars appear and grow in the Universe.

\acknowledgements

H.C. and N. G. thank the anonymous referee for the valuable suggestions that help improve this work. H.C. and N. G. also thank Anna-Christina Eilers and James Bolton for valuable comments.
This work was supported by the NASA ATP grant NNX17AK65G and NASA FINESST grant NNH19ZDA005K. This project is carried out on the Midway cluster at the University of Chicago Research Computing Center.

\bibliographystyle{apj}
\bibliography{main}

\begin{thebibliography}{23}
\expandafter\ifx\csname natexlab\endcsname\relax\def\natexlab#1{#1}\fi

\bibitem[{{Bolton} \& {Haehnelt}(2007)}]{bolton2007}
{Bolton}, J.~S. \& {Haehnelt}, M.~G. 2007, \mnras, 374, 493

\bibitem[{{Bosman} {et~al.}(2020){Bosman}, {{\v{D}}urov{\v{c}}{\'\i}kov{\'a}},
  {Davies}, \& {Eilers}}]{bosman2020}
{Bosman}, S.~E.~I., {{\v{D}}urov{\v{c}}{\'\i}kov{\'a}}, D., {Davies}, F.~B., \&
  {Eilers}, A.~C. 2020, arXiv e-prints, arXiv:2006.10744

\bibitem[{{Carilli} {et~al.}(2010){Carilli}, {Wang}, {Fan}, {Walter}, {Kurk},
  {Riechers}, {Wagg}, {Hennawi}, {Jiang}, {Menten}, {Bertoldi}, {Strauss}, \&
  {Cox}}]{carilli2010}
{Carilli}, C.~L., {Wang}, R., {Fan}, X., {Walter}, F., {Kurk}, J., {Riechers},
  D., {Wagg}, J., {Hennawi}, J., {Jiang}, L., {Menten}, K.~M., {Bertoldi}, F.,
  {Strauss}, M.~A., \& {Cox}, P. 2010, \apj, 714, 834

\bibitem[{{Chen} \& {Gnedin}(2021)}]{chen2021}
{Chen}, H. \& {Gnedin}, N.~Y. 2021, \apj, 911, 60

\bibitem[{{Croft} {et~al.}(1998){Croft}, {Weinberg}, {Katz}, \&
  {Hernquist}}]{croft1998}
{Croft}, R. A.~C., {Weinberg}, D.~H., {Katz}, N., \& {Hernquist}, L. 1998,
  \apj, 495, 44

\bibitem[{{Davies} {et~al.}(2018){Davies}, {Hennawi}, {Ba{\~n}ados}, {Simcoe},
  {Decarli}, {Fan}, {Farina}, {Mazzucchelli}, {Rix}, {Venemans}, {Walter},
  {Wang}, \& {Yang}}]{davies2018}
{Davies}, F.~B., {Hennawi}, J.~F., {Ba{\~n}ados}, E., {Simcoe}, R.~A.,
  {Decarli}, R., {Fan}, X., {Farina}, E.~P., {Mazzucchelli}, C., {Rix}, H.-W.,
  {Venemans}, B.~P., {Walter}, F., {Wang}, F., \& {Yang}, J. 2018, \apj, 864,
  143

\bibitem[{{Davies} {et~al.}(2020){Davies}, {Hennawi}, \& {Eilers}}]{davies2020}
{Davies}, F.~B., {Hennawi}, J.~F., \& {Eilers}, A.-C. 2020, \mnras, 493, 1330

\bibitem[{{Eilers} {et~al.}(2017){Eilers}, {Davies}, {Hennawi}, {Prochaska},
  {Luki{\'c}}, \& {Mazzucchelli}}]{eilers2017}
{Eilers}, A.-C., {Davies}, F.~B., {Hennawi}, J.~F., {Prochaska}, J.~X.,
  {Luki{\'c}}, Z., \& {Mazzucchelli}, C. 2017, \apj, 840, 24

\bibitem[{{Fan} {et~al.}(2006){Fan}, {Strauss}, {Becker}, {White}, {Gunn},
  {Knapp}, {Richards}, {Schneider}, {Brinkmann}, \& {Fukugita}}]{fan2006}
{Fan}, X., {Strauss}, M.~A., {Becker}, R.~H., {White}, R.~L., {Gunn}, J.~E.,
  {Knapp}, G.~R., {Richards}, G.~T., {Schneider}, D.~P., {Brinkmann}, J., \&
  {Fukugita}, M. 2006, \aj, 132, 117

\bibitem[{{Fukugita} {et~al.}(1998){Fukugita}, {Hogan}, \&
  {Peebles}}]{fukugita1998}
{Fukugita}, M., {Hogan}, C.~J., \& {Peebles}, P.~J.~E. 1998, \apj, 503, 518

\bibitem[{Jiang {et~al.}(2016)Jiang, McGreer, Fan, Strauss, Ba{\~{n}}ados,
  Becker, Bian, Farnsworth, Shen, Wang, Wang, Wang, White, Wu, Wu, Yang, \&
  Yang}]{jiang2016}
Jiang, L., McGreer, I.~D., Fan, X., Strauss, M.~A., Ba{\~{n}}ados, E., Becker,
  R.~H., Bian, F., Farnsworth, K., Shen, Y., Wang, F., Wang, R., Wang, S.,
  White, R.~L., Wu, J., Wu, X.-B., Yang, J., \& Yang, Q. 2016, The
  Astrophysical Journal, 833, 222

\bibitem[{{Keating} {et~al.}(2015){Keating}, {Haehnelt}, {Cantalupo}, \&
  {Puchwein}}]{keating2015}
{Keating}, L.~C., {Haehnelt}, M.~G., {Cantalupo}, S., \& {Puchwein}, E. 2015,
  \mnras, 454, 681

\bibitem[{{Kirkman} {et~al.}(2005){Kirkman}, {Tytler}, {Suzuki}, {Melis},
  {Hollywood}, {James}, {So}, {Lubin}, {Jena}, {Norman}, \&
  {Paschos}}]{kirkman2005}
{Kirkman}, D., {Tytler}, D., {Suzuki}, N., {Melis}, C., {Hollywood}, S.,
  {James}, K., {So}, G., {Lubin}, D., {Jena}, T., {Norman}, M.~L., \&
  {Paschos}, P. 2005, \mnras, 360, 1373

\bibitem[{{Lusso} {et~al.}(2015){Lusso}, {Worseck}, {Hennawi}, {Prochaska},
  {Vignali}, {Stern}, \& {O'Meara}}]{lusso2015}
{Lusso}, E., {Worseck}, G., {Hennawi}, J.~F., {Prochaska}, J.~X., {Vignali},
  C., {Stern}, J., \& {O'Meara}, J.~M. 2015, \mnras, 449, 4204

\bibitem[{{Matsuoka} {et~al.}(2018){Matsuoka}, {Strauss}, {Kashikawa}, {Onoue},
  {Iwasawa}, {Tang}, {Lee}, {Imanishi}, {Nagao}, {Akiyama}, {Asami}, {Bosch},
  {Furusawa}, {Goto}, {Gunn}, {Harikane}, {Ikeda}, {Izumi}, {Kawaguchi},
  {Kato}, {Kikuta}, {Kohno}, {Komiyama}, {Lupton}, {Minezaki}, {Miyazaki},
  {Murayama}, {Niida}, {Nishizawa}, {Noboriguchi}, {Oguri}, {Ono}, {Ouchi},
  {Price}, {Sameshima}, {Schulze}, {Shirakata}, {Silverman}, {Sugiyama},
  {Tait}, {Takada}, {Takata}, {Tanaka}, {Toba}, {Utsumi}, {Wang}, \&
  {Yamashita}}]{matsuka2018}
{Matsuoka}, Y., {Strauss}, M.~A., {Kashikawa}, N., {Onoue}, M., {Iwasawa}, K.,
  {Tang}, J.-J., {Lee}, C.-H., {Imanishi}, M., {Nagao}, T., {Akiyama}, M.,
  {Asami}, N., {Bosch}, J., {Furusawa}, H., {Goto}, T., {Gunn}, J.~E.,
  {Harikane}, Y., {Ikeda}, H., {Izumi}, T., {Kawaguchi}, T., {Kato}, N.,
  {Kikuta}, S., {Kohno}, K., {Komiyama}, Y., {Lupton}, R.~H., {Minezaki}, T.,
  {Miyazaki}, S., {Murayama}, H., {Niida}, M., {Nishizawa}, A.~J.,
  {Noboriguchi}, A., {Oguri}, M., {Ono}, Y., {Ouchi}, M., {Price}, P.~A.,
  {Sameshima}, H., {Schulze}, A., {Shirakata}, H., {Silverman}, J.~D.,
  {Sugiyama}, N., {Tait}, P.~J., {Takada}, M., {Takata}, T., {Tanaka}, M.,
  {Toba}, Y., {Utsumi}, Y., {Wang}, S.-Y., \& {Yamashita}, T. 2018, \apj, 869,
  150

\bibitem[{{McDonald} {et~al.}(2000){McDonald}, {Miralda-Escud{\'e}}, {Rauch},
  {Sargent}, {Barlow}, {Cen}, \& {Ostriker}}]{mcdonald2000}
{McDonald}, P., {Miralda-Escud{\'e}}, J., {Rauch}, M., {Sargent}, W. L.~W.,
  {Barlow}, T.~A., {Cen}, R., \& {Ostriker}, J.~P. 2000, \apj, 543, 1

\bibitem[{{McQuinn}(2016)}]{mcquinn2016}
{McQuinn}, M. 2016, \araa, 54, 313

\bibitem[{{Onoue} {et~al.}(2017){Onoue}, {Kashikawa}, {Willott}, {Hibon}, {Im},
  {Furusawa}, {Harikane}, {Imanishi}, {Ishikawa}, {Kikuta}, {Matsuoka},
  {Nagao}, {Niino}, {Ono}, {Ouchi}, {Tanaka}, {Tang}, {Toshikawa}, \&
  {Uchiyama}}]{onoue2017}
{Onoue}, M., {Kashikawa}, N., {Willott}, C.~J., {Hibon}, P., {Im}, M.,
  {Furusawa}, H., {Harikane}, Y., {Imanishi}, M., {Ishikawa}, S., {Kikuta}, S.,
  {Matsuoka}, Y., {Nagao}, T., {Niino}, Y., {Ono}, Y., {Ouchi}, M., {Tanaka},
  M., {Tang}, J.-J., {Toshikawa}, J., \& {Uchiyama}, H. 2017, \apjl, 847, L15

\bibitem[{{Padmanabhan} {et~al.}(2014){Padmanabhan}, {Choudhury}, \&
  {Srianand}}]{padmanabhan2014}
{Padmanabhan}, H., {Choudhury}, T.~R., \& {Srianand}, R. 2014, \mnras, 443,
  3761

\bibitem[{{Rauch} {et~al.}(1997){Rauch}, {Miralda-Escud{\'e}}, {Sargent},
  {Barlow}, {Weinberg}, {Hernquist}, {Katz}, {Cen}, \& {Ostriker}}]{rauch1997}
{Rauch}, M., {Miralda-Escud{\'e}}, J., {Sargent}, W. L.~W., {Barlow}, T.~A.,
  {Weinberg}, D.~H., {Hernquist}, L., {Katz}, N., {Cen}, R., \& {Ostriker},
  J.~P. 1997, \apj, 489, 7

\bibitem[{{Seljak} {et~al.}(2005){Seljak}, {Makarov}, {McDonald}, {Anderson},
  {Bahcall}, {Brinkmann}, {Burles}, {Cen}, {Doi}, {Gunn}, {Ivezi{\'c}}, {Kent},
  {Loveday}, {Lupton}, {Munn}, {Nichol}, {Ostriker}, {Schlegel}, {Schneider},
  {Tegmark}, {Berk}, {Weinberg}, \& {York}}]{seljak2005}
{Seljak}, U., {Makarov}, A., {McDonald}, P., {Anderson}, S.~F., {Bahcall},
  N.~A., {Brinkmann}, J., {Burles}, S., {Cen}, R., {Doi}, M., {Gunn}, J.~E.,
  {Ivezi{\'c}}, {\v{Z}}., {Kent}, S., {Loveday}, J., {Lupton}, R.~H., {Munn},
  J.~A., {Nichol}, R.~C., {Ostriker}, J.~P., {Schlegel}, D.~J., {Schneider},
  D.~P., {Tegmark}, M., {Berk}, D.~E., {Weinberg}, D.~H., \& {York}, D.~G.
  2005, \prd, 71, 103515

\bibitem[{{Viel} {et~al.}(2005){Viel}, {Lesgourgues}, {Haehnelt}, {Matarrese},
  \& {Riotto}}]{viel2005}
{Viel}, M., {Lesgourgues}, J., {Haehnelt}, M.~G., {Matarrese}, S., \& {Riotto},
  A. 2005, \prd, 71, 063534

\bibitem[{{Willott} {et~al.}(2010){Willott}, {Delorme}, {Reyl{\'e}}, {Albert},
  {Bergeron}, {Crampton}, {Delfosse}, {Forveille}, {Hutchings}, {McLure},
  {Omont}, \& {Schade}}]{willott2010}
{Willott}, C.~J., {Delorme}, P., {Reyl{\'e}}, C., {Albert}, L., {Bergeron}, J.,
  {Crampton}, D., {Delfosse}, X., {Forveille}, T., {Hutchings}, J.~B.,
  {McLure}, R.~J., {Omont}, A., \& {Schade}, D. 2010, \aj, 139, 906

\end{thebibliography}

\end{document}